# A versatile three-dimensional traction force microscopy framework for uncovering the mechanics of bio-adhesion

Yingwei Hou[1] (Yingwei.hou@qmul.ac.uk), Fusheng Wang[2] (fswang@nwpu.edu.cn), Tao Liu[1*] (Tao.Liu@qmul.ac.uk)

[1]School of Engineering and Materials Science, Queen Mary University of London, E1 4NS, UK

[2]School of Mechanics and Transportation Engineering, Northwestern Polytechnical University, 710129 Xi'an, P.R. China.

[*]Corresponding author: Tao.Liu@qmul.ac.uk; fswang@nwpu.edu.cn

**Data Availability Statement**
All data supporting the findings of this study are included in the manuscript and Supplementary Materials. Additional data and code are available from the corresponding author upon reasonable request.

**Funding Statement**
This work was supported by the Leverhulme Trust Research Grant Scheme, UK (No. RPG-2020-235).

**Conflict of Interest Disclosure**
The authors declare no conflicts of interest.

**Ethics Approval Statement**
Marine mussel plaques were harvested from live mussels by gently cutting threads. All procedures involving live marine invertebrates were conducted in accordance with institutional and ethical guidelines for the care and use of marine organisms. No procedures involving vertebrate animals or human subjects were performed.

**Abstract**

This study presents a traction force microscopy framework for quantifying three-dimensional interfacial forces during bio-adhesion by integrating in-situ stereo-digital image correlation with finite element (FE) simulation. The method enables accurate measurement of microscale displacements and force distributions in both dry and wet environments, addressing limitations of conventional microscopy techniques related to limited measurement scales, restricted fields of view, and surface disturbance from contact or fluorescence. A model was developed to guide the design of a polydimethylsiloxane substrate with optimized Young's modulus and transparency to improve accuracy. System accuracy was validated through FE simulations and steel ball compression experiments. The framework was applied to marine mussel plaque adhesion under 15° directional tension to characterize interfacial force distributions. Sensitivity analyses examined the effects of Poisson's ratio, Young's modulus, and constitutive models on the results. This approach offers a versatile platform for investigating interfacial mechanics in adhesives, with broad relevance to bioengineering applications.

**Keywords:** Traction Force measurement; Traction force microscopy; Digital image correlation; Interfacial mechanics; Wet adhesion; Finite element

## 1. Introduction

Traction forces, e.g. cohesive forces and contact forces, play an important role in wide ranges of applications such as adhesives (*1*), coatings (*2*) and medical applications (*3*). The mechanism of traction forces is crucial for optimizing material performance by improving energy efficiency, ensuring structure integrity as well as developing material design. In complex systems, understanding force distribution across the interface between two contact surfaces is important for developing durable and efficient load-bearing structures.

Accurate measurement of force distribution at interfaces is essential to investigate the mechanisms of traction force. Several characterising techniques have been employed to measure traction force. For example, atomic force microscopy (AFM) is used to measure adhesive forces by detecting forces between a tip and a sample as the tip scans the sample surface (*4*). AFM has a high-resolution at atomic scale, however, its scanning speed is slow to avoid compromising the resolution (*5*). Furthermore, fluorescence microscopy (FM) is used to measure traction forces by tracking the behavior of fluorescently labelled molecules or particles near surfaces (*6*). FM is subject to photobleaching, where fluorophores irreversibly degrade under prolonged illumination (*7*), and phototoxicity caused by reactive oxygen species, which can alter cellular physiology (*8*). These effects are particularly problematic in live imaging and therefore make such techniques less suitable for experiments involving living animals or delicate biological interfaces (*9*). Traction force microscopy (TFM) is a widely used technique for quantifying forces exerted by cells on the surface of substrates. The traction forces measured at the cell–substrate interface typically represent internal forces exerted by adherent cells (*10*). TFM is generally applied to assess cellular deformations at limited spatial scales, ranging from nanometre to submicron (*11–13*), and primarily provides a two-dimensional view of traction forces. As TFM commonly relies on confocal fluorescence microscopy for displacement measurements (*14–16*), it is also subject to the inherent limitations of fluorescence microscopy, including the above-mentioned photobleaching and phototoxicity.

Recent advancements in multidimensional traction force microscopy have significantly expanded our ability to quantify mechanical transduction in 3D environments, as reviewed by Cheung et al (*17*). Pioneering methods utilizing confocal fluorescence microscopy, such as those by Legant et al (*18*). and Plotnikov (*19*) et al., have enabled the mapping of traction forces with sub-micron resolution, revealing detailed rotational moments and focal adhesion dynamics at the single-cell level. Subsequent optimizations by Holenstein et al. (*20*) and Bergert et al. (*21*) further refined tracking accuracy and extended these capabilities to analyzing adhesion-independent migration. More recently, Lee et al. (*22*) introduced Refractive-Index TFM (RI-TFM), a label-free tomographic approach that overcomes the limitations of phototoxicity and slow acquisition speeds inherent to fluorescence-based scanning.

However, despite their high resolution, these microscopy-based techniques are fundamentally constrained by the field of view (FOV) and working distance of high-magnification objectives (Table 1). Furthermore, these techniques are predominantly optimized for assessing cellular deformations which typically remain below 2 μm (*23–26*). Such limitations make them unsuitable for characterizing meso-scale or tissue-level mechanics, particularly when investigating large, thick samples or applying in situ external loading. To address this gap, we present a stereo-DIC-based framework designed for macroscopic interfacial force mapping. Unlike confocal methods, our approach creates a versatile, phototoxicity-free platform capable of measuring surface deformations across centimeter-scale, unrestricted by the spatial constraints of standard microscopy.

Table 1. Comparison of representative TFM methods and the present framework

| **Authors** | **Scale (resolution)** | **Imaging & Requirements** | **Applicability & Limitations** |
|---|---|---|---|
| Legant et al.(*18*) | **Single Cell** (Voxel dimensions: 0.19 × 0.19 × 0.5 μm) | **Confocal Microscopy** Requires hydrogel with fluorescent beads; Fluorescence labelling | **3D sub-cellular force mapping** Limited by phototoxicity and slow z-stack acquisition; Limited to linear elastic substrate |
| Plotnikov et al. (*19*) | **Single Cell** (Window size: 0.94 × 0.94 μm) | **Confocal Microscopy** Requires high-density; Fluorescent beads | **2D traction at focal adhesions** Restricted to linear elastic substrate |
| Holenstein et al.(*20*) | **Single Cell** (Voxel dimensions: 0.06 × 0.06 ×0.25 μm) | **Confocal Microscopy** Optimized bead density and optical flow tracking | **High-accuracy tracking** Improves resolution for small focal adhesions; Limited to linear elastic substrate |
| Bergert et al. (*21*) | **Single Cell** | **Confocal Microscopy** Fluorescence imaging of beads; printed fiducial arrays | **3D traction mapping of cell migration** Require image stacking and reconstruction; Computationally expensive |
| Lee et al.(*22*) | **Single Cell** (Voxel dimensions: 0.18 × 0.18 ×0.18 μm) | **Refractive-Index Tomography** Label-free (Refractive Index) | **High-speed 3D mapping** Suitable for capturing rapid dynamics; Require image stacking and reconstruction; Limited to single-cell FOV |
| Bergert et al. (*27*) | **Single Cell** | **Confocal Microscopy** Nanodrop-printed fiducial arrays; Confocal fluorescence imaging | **Reference-Free TFM** Capturing dynamic forces where a reference state is impossible; Computationally expensive |
| This work | **Tissue scale** Pixel size: 15 ×15 um | **Stereo-DIC (Cameras)** Standard cameras with surface speckle/pigment; No fluorescence or microscopy required | **Macro-scale Interfacial mapping** No phototoxicity; Suitable for large samples (unrestricted by microscope FOV); Enable application of external loading |

Among the developing approaches, Li et al. (*28*) presented a light field microscopy-based technique that enables dynamic imaging of 3D traction forces on polydimethylsiloxane (PDMS) substrates, based on an iterative displacement to stress inversion algorithm. While their method provides reconstruction of 3D traction stresses, the assumption that PDMS substrates behave as linear elastic solids may compromise the accuracy of traction force measurements, particularly in regions of local deformation. Since PDMS is known to exhibit pronounced hyperelastic behavior (*29*, *30*), accurately modeling its nonlinear mechanical response is essential for reliable force quantification. In addition, the influence of material parameters, such as Young's modulus and Poisson's ratio, on the accuracy of force reconstruction has not yet been systematically investigated. In this work, we address these open questions by integrating experimentally measured displacements into a finite element framework incorporating a hyperelastic substrate model, and by quantifying the influence of substrate properties on interfacial force measurements. This approach offers an experimentally simpler and versatile platform for displacement measurements using transparent layered polymeric substrates. Unlike confocal microscopy-based methods, which often restrict the dimensions of the sample size due to limited field of view and working distance, this setup accommodates a wider range of sample geometries while maintaining high-resolution surface deformation tracking. Moreover, it is particularly suited for in situ applications involving external loading, such as adhesion testing under wet conditions, which remains challenging for existing optical based 3D traction measurement techniques.

This paper aims to bridge this gap by providing an in-situ technique that enables the real-time measurement of user-specific displacements and 3D traction force distributions at the interface. As an illustration of this technique, the interfacial traction force of marine mussel plaques anchored to wet substrates under directional tension is measured. The technique is applicable to any interfacial interactions for substrates with different stiffness and not limited to adhesion-detachment related deformation at the interface.

## 2. Results and discussion

### 2.1 Experimental design

The measurement of 3D displacement field and force distribution at an interface was conducted using an in-situ stereo-digital image correlation (SDIC) method. SDIC is a non-contact optical technique, and its setup and principle are illustrated in Figure 1. The setup (Figure 1a) consists of a substrate and charge-coupled device (CCD) cameras. In this study, PDMS made of Sylgard 184 silicone elastomer (The Dow Chemical Company, Michigan, USA) was used to create the deformable substrate for measuring the mechanical responses at the interfaces. PDMS was chosen for this method as it is optically transparent, non-toxic, and highly deformable (*31*).

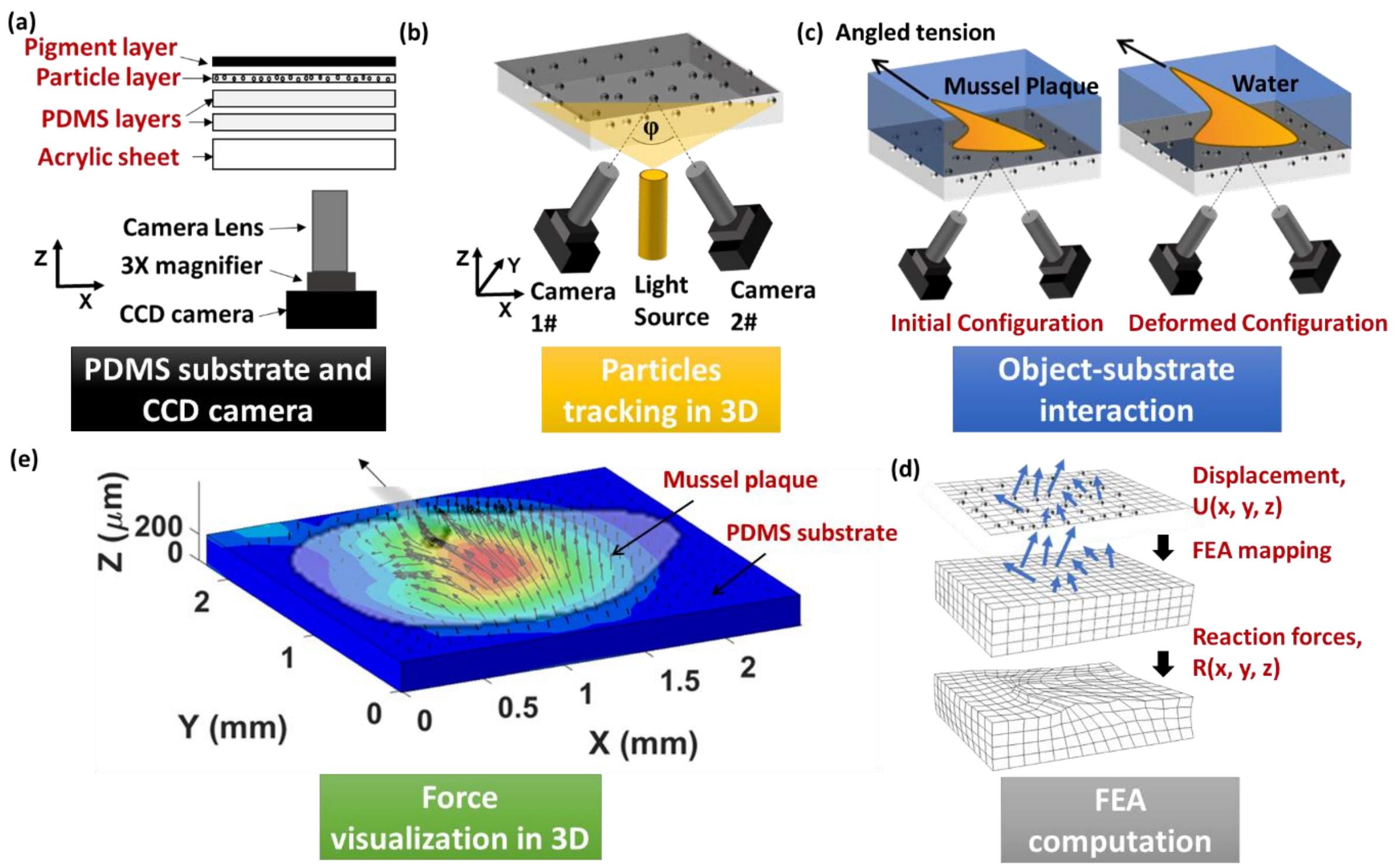

Figure 1. The schematic of the SDIC technique for measuring 3D interfacial force in wet environment: (a) the PDMS substrate consists of four layers and is supported on an acrylic panel; (b) two cameras with an angle φ underneath the substrate; (c) an object under directional tension in water while the two cameras measuring the substrate's deformation at interface; (d) importing the measured 3D displacement U(x, y, z) into the finite element (FE) model of the substrate via coordinate mapping and computing 3D reaction force R(x, y, z) at the interface; (e) the visualization of the 3D traction force at the interface between the object and substrate.

Four PDMS layers were coated separately on an acrylic substrate. The four layers from the bottom to top were two layers pure PDMS, a particle layer and a pigment layer, shown in Figure 2a. The first two layers made of pure PDMS were used as a deformable substrate. The third PDMS layer contained randomly distributed particles (ZnS:Cu (*32*)) forming a speckle pattern with particle sizes of 3.0 ± 1.2 pixels (45 ± 18 μm) and areal coverage between 20% and 40%, making it suitable for digital image correlation (DIC) measurement (*33*, *34*). The black pigment layer as the fourth layer was used to alleviate the scattering light by water at the wet interface and increase the contrast of the white micro particles relative to surrounding PDMS. The third and fourth layers collaborated to form a uniform, high-contrast, and randomly distributed speckle pattern on the top surface of the PDMS substrate (Figure 2b). The overall thickness of the four PDMS substrate was 0.3 mm. The PDMS substrate preparation and spin coating process were detailed in Supplementary Materials S.1.

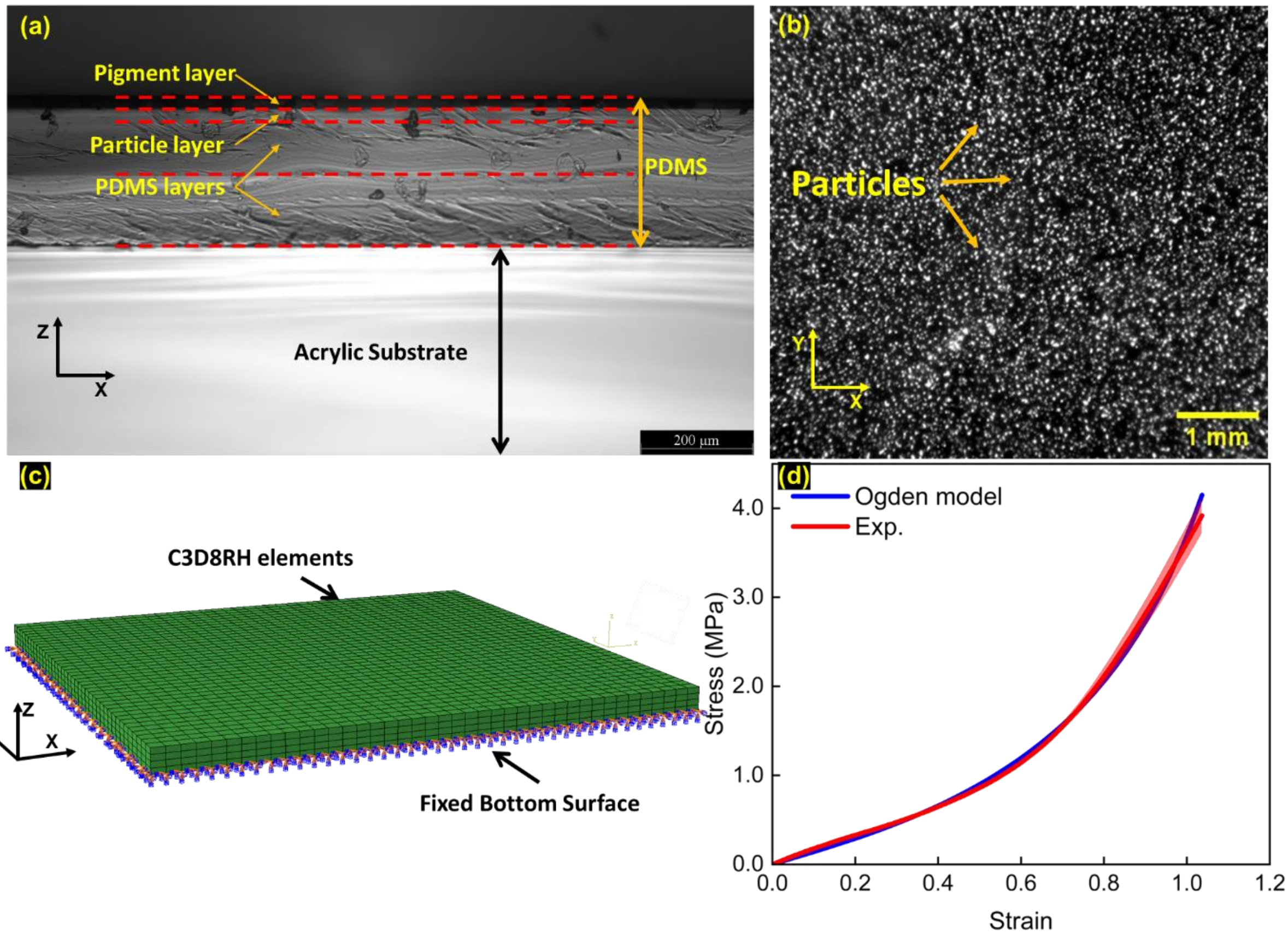

Figure 2. (a) The optical microscope image of the PDMS substrate cross-section, which consists of four layers from bottom to top: two layers pure PDMS, a particle layer and a pigment layer; (b) an image taken from a CCD camera showing speckle patterns consisting of randomly distributed particles and surrounding PDMS. (c) the FE model of the PDMS substrate using

commercial FE solver Abaqus/standard. (d) The stress-strain curve of the PDMS substrate obtained by uniaxial tensile test and curve fitting using Ogden model.

Two synchronized cameras (Figure 1b) were employed to capture images of the substrate from two different viewpoints, forming a stereo pair. The stereo images were analysed using DIC algorithms embedded in a processing software (DICe (*35*), an open-source software tool developed by Sandia National Laboratories for performing DIC to identify matching points on the speckle pattern across the stereo pair. The 3D spatial position of a point was determined through triangulation (*36*) which relied on the intersection of rays from the cameras. This approach utilized camera parameters (e.g., focal length, location) and the pixel coordinates of the corresponding point in both images, obtained by a prior calibration process (*37*).

Using the reconstructed 3D speckle pattern, DICe software calculates 3D displacement vectors by tracking the relative movement of points between the initial (undeformed) and deformed configurations, caused by interactions at the interface (e.g., mussel plaques attached to a substrate, as shown in Figure 1c). The measured displacements include components along the X and Y axes, representing in-plane directions, and the Z axis, representing the out-of-plane direction.

The displacement vectors at each point across the interface were mapped onto the nodes of the FE model (ABAQUS®/Standard) of the PDMS substrate as displacement boundary conditions (Figure 1d). To ensure accurate boundary condition application, the finite element mesh was constructed directly based on the DIC analysis grid. Specifically, the spatial coordinates of the FE nodes were defined to coincide exactly with the central points of the DIC subsets. This establishes a one-to-one mapping between the experimental measurements and numerical models, i.e., no interpolation is required during the mapping. Consequently, the 3D displacement vectors (X, Y, and Z components) measured at each DIC subset centre were directly assigned as displacement boundary conditions to the corresponding FE node. The FE model was assigned the same thickness as the substrate used in the experiment. The PDMS substrate was modelled using reduced hybrid 3D 8-node elements (C3D8RH in ABAQUS notation), shown in Figure 2c. The nodes at the bottom surface were constrained to prevent any translation or rotation movement. The mesh density of the FE model was approximately 475 elements/mm$^3$, which is consistent with experimental DIC mesh resolution. A mesh density study confirmed that the results had converged: no significant changes upon further mesh refinement, demonstrating that the selected mesh was sufficient for accurate and reliable analysis.

The PDMS exhibits a nonlinear hyperelastic behaviour which was modelled using the compressible Ogden model (*38*) in this study, as detailed below:

$$W(\lambda_1, \lambda_2, \lambda_3) = \sum_{p=1}^{N} \frac{\mu_p}{\alpha_p} \left( \lambda_1^{\alpha_p} + \lambda_2^{\alpha_p} + \lambda_3^{\alpha_p} - 3 \right) + \frac{1}{D} (J - 1)^2 \quad (1)$$

where $W$ is strain energy density; $\lambda_1, \lambda_2, \lambda_3$ are the principal stretch ratios (eigenvalues of the deformation gradient); $\mu_p$ and $\alpha_p$ material constraints with the initial shear modulus calculated as $\sum_{p=1}^{N} \mu_p$; $N$ the order of energy potential; $J$ the volume ratio, $J = \lambda_1 \lambda_2 \lambda_3$; $D$ the material incompressibility parameter which can be related to initial bulk modulus $K = 2/D$. It was found that the Ogden model best fit to the uniaxial test data (ASTM-D412 (*39*)) of pure PDMS when $N = 2$, shown in Figure 2d. Table 2 summarises the material constants used in the Ogden model. It is worth mentioning that uniaxial test (ASTM D882 (*40*)) was also conducted on the PDMS containing particles and pigment, which was cut directly from the substrate. ASTM D412 was used for pure PDMS as an elastomer with large-strain behaviour (>100%), while ASTM D882 was more suitable for the particle-filled PDMS as a thin film with lower strain (~25%), close to the strain ranges used in this study. The results indicated that addition of particles and pigment had a negligible impact on the mechanical behaviour, as the stress-strain curve is almost identical to that of pure PDMS, see Supplementary Materials S.2.

Table 2. Material constants employed in the Ogden model

| $\mu_1$ (MPa) | $\mu_2$ (MPa) | $\alpha_1$ | $\alpha_2$ | $D_1$ | Young's modulus (MPa) | Poisson's Ratio |
|---|---|---|---|---|---|---|
| 4.71E-04 | 0.47 | 14.79 | 4.84 | 0.44 | 1.38 | 0.45 |

The Poisson's ratio of the PDMS substrate was experimentally measured using DIC-based uniaxial tensile testing on thin films directly cut from the same formulation used in this study, yielding a value of $\nu$=0.45 ± 0.03, consistent with literature values for cross-linked PDMS of similar composition (*41–43*). The detailed procedures and results are provided in Supplementary Materials S3.

To verify that the speckle treatment did not bias adhesion measurements, control experiments were conducted on pristine and speckled PDMS substrates. Contact angle measurements showed no significant difference between the two surfaces (109.2° ± 4.7° vs. 110.2° ± 6.1°), indicating that the addition of ZnS:Cu particles and black pigments did not alter the surface energy. Adhesion performance was evaluated using a standard ASTM D903 peeling test with a commercial bonding tape to ensure reproducibility. The speckled PDMS exhibited a slightly higher peeling energy, which may be resulted from enhanced microscale mechanical interlocking due to the particles modified the microscale topography (see Supplementary Materials S.4). These results indicate that while surface chemistry remains consistent, the topographical changes induced by the speckle pattern can influence mechanical adhesion. Therefore, further optimization of particle density and size is required to mitigate the mechanical effects of particulate additives on adhesion in future applications.

Based on the FE model of the PDMS substrate, reaction forces (nodal forces) at the top surface of the substrate were computed in ABAQUS®/Standard. As an example, the method was applied to measure and visualize the 3D reaction force at the interface between a mussel plaque and the PDMS substrate (Figure 1e).

## 2.2 Design of the substrate

The material selection and thickness of the substrate's deformable layers are crucial for measurement accuracy. This study introduces a quantitative model to determine the suitable substrate material and thickness, based on the Winkler Spring model that treats the substrate as an isotropic elastic-foundation composed of independent vertical springs, as shown in Figure 3a.

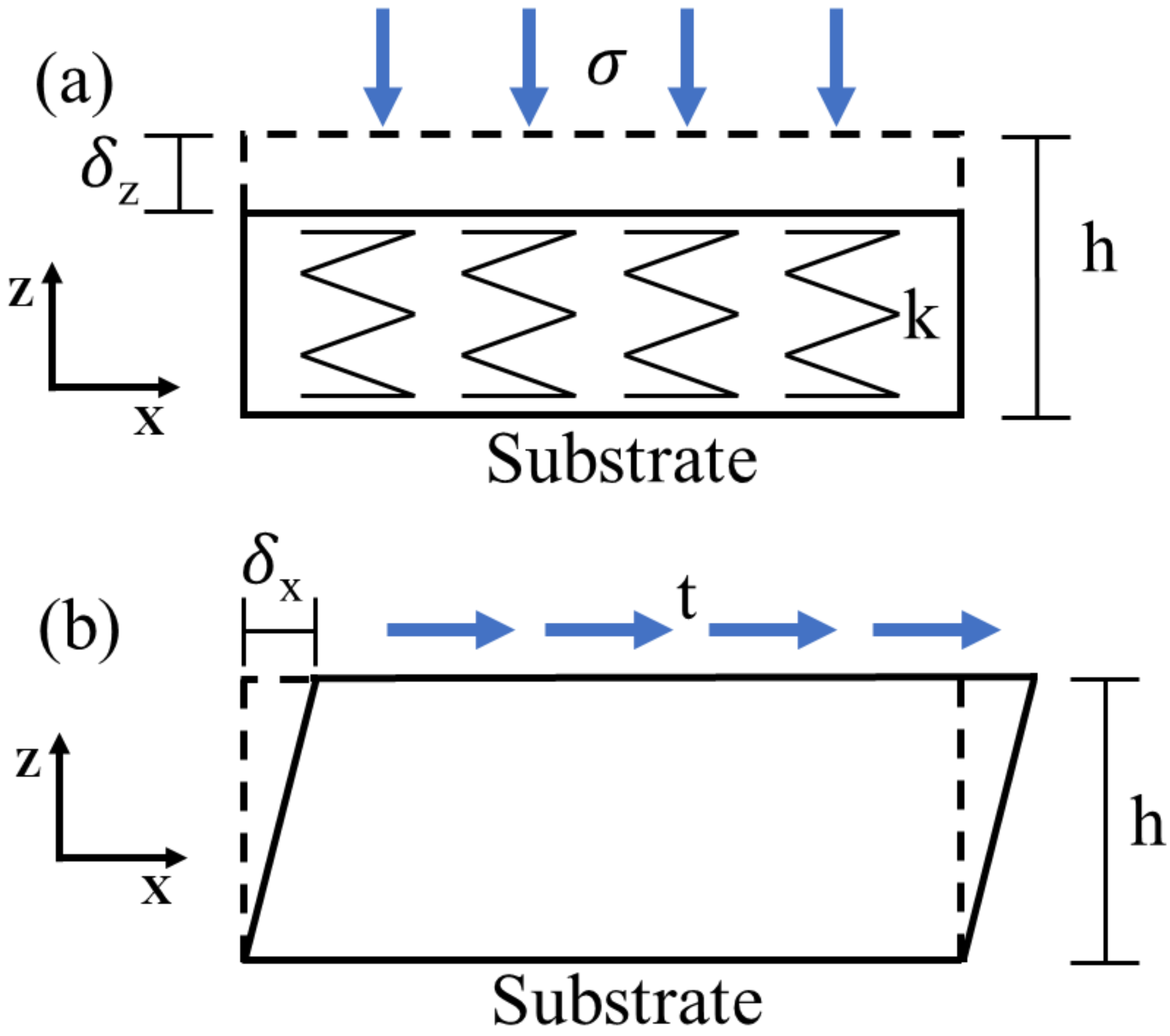

Figure 3. Schematics of (a) the Winkler Spring model under vertical contact stress $\sigma$, and (b) the deformed configuration of the substrate under shear traction $t$.

The elastic stiffness of a Winkler foundation $k$ can be related to the vertical contact stress σ and the local vertical deformation $\delta_z$ via

$$k = \sigma \,/\, \delta_z \tag{2}$$

The value of elastic stiffness $k$ can also be related to Young's modulus $E$ and thickness $h$ of the substrate (*44*), i.e.,

$$k = \frac{E}{h} \tag{3}$$

The classical Winkler foundation model assumes that the substrate reacts locally to the applied load, neglecting the lateral coupling between adjacent points. This simplification neglects lateral constraint and Poisson coupling, leading to discrepancies for finite-thickness elastic layers. Based on the Timoshenko solution (*45*), the stiffness of a finite-thickness layer, $k_T$ can be written as

$$k_T = \frac{E}{h \times (1 - v^2)} \tag{3}$$

To evaluate this approximation under the experimental conditions of this study, FE simulations were performed using a 3D elastic substrate (E = 1.7 MPa, ν = 0.3–0.45, h = 0.3 mm) and subjected to uniform z-direction pressure. The effective stiffness obtained from simulation, $k_{FE} = {}^{\sigma}/_{\delta_z}$, was compared with the analytical expression Equation (3), and a modifying parameter $\alpha(v)$ was defined as:

$$\alpha(v) = \frac{E/(h \times (1 - v^2))}{k_{FE}} = -12.50 \times v^2 + 6.35 \times v \tag{4}$$

Polynomial fitting of $\alpha(v)$ yielded a correlation $R^2 = 1.0$, confirming that the modified Winkler model reproduces the FE-derived stiffness precisely across the investigated Poisson-ratio range (see Supplementary Materials S4).

By combining $\alpha(v)$ with the Poisson-dependent term from the analytical model $k_T$, a unified coefficient $\theta(v) = \frac{1}{\alpha(v) \times (1 - v^2)}$ was defined, giving the modified Winkler foundation model expression:

$$k = \frac{E}{h} \times \theta(v) = \frac{E}{h} \times \frac{1}{(1 - v^2) \times (-12.50 \times v^2 + 6.35 \times v)} \tag{3}$$

This modified form consolidates the effects of lateral constraint and Poisson coupling into a single function of ν, providing an accurate and simple representation of the finite-thickness elastic response validated by the FE simulations as illustrated in Supplementary Materials S5.

From Equations (1) and (2), the ratio ($\psi_z$) of noise ($\eta_z$) to measured deformation ($\delta_z$) in vertical direction (Z direction) can be related to Young's modulus $E$ and thickness $h$ of the substrate, i.e.,

$$\Psi_z = {}^{\eta_z}/_{\delta_z} = \frac{\eta_z \times k}{\sigma} = \frac{\eta_z \times E \ \times \theta(v)}{\sigma \times h} \tag{5}$$

where $\eta_z$ denotes the measurement noise in Z direction.

Consider a shear traction ($t$) that is applied parallelly on the top surface and causes deformation $\delta_x$ along the X-axis (see Figure 3b), the ratio of the noise ($\eta_x$) to measured deformation $\psi_x$ can be calculated as

$$\Psi_x = {}^{\eta_x}/_{\delta_x} = \frac{\eta_x \times \ G}{t \times h} = \frac{\eta_x \times E}{t \times h \times 2(1+v)} \tag{6}$$

where $G$ and $v$ are the shear modulus and Poisson's ratio, respectively. Equations (5) and (6) can be combined to describe the general noise to measurement ratio ($\psi$) in three directions (X-Y-Z) as below:

$$\Psi = {}^{\eta}/_{\delta} = \frac{\sqrt{{\eta_x}^2 + {\eta_y}^2 + {\eta_z}^2}}{\sqrt{{\delta_x}^2 + {\delta_y}^2 + {\delta_z}^2}} \tag{7}$$

Assuming the in-plane deformation along the Y-axis ($\delta_y$) is equal to $\delta_x$, and $\eta_{max} = max(\eta_x, \eta_y, \eta_z)$, Equation (7) can be written as follows:

$$\Psi = \frac{\sqrt{{\eta_x}^2 + {\eta_y}^2 + {\eta_z}^2}}{\sqrt{{\delta_x}^2 + {\delta_y}^2 + {\delta_z}^2}} \leq \frac{\sqrt{3{\eta_{max}}^2}}{\sqrt{2 \times \frac{2(1+v) \times t \times h}{E}^2 + \left(\frac{\sigma \times h}{E \times \theta(v)}\right)^2}}$$

$$= \frac{E \times \eta_{max}}{h} \times \frac{\sqrt{3}}{\sqrt{2(2(1+v) \times t)^2 + (\sigma/\theta(v))^2}} \quad (8)$$

Equation (8) suggests that the value of $\psi$ can be reduced by using a soft substrate (low Youngs's modulus *E*) and increasing the thickness *h* of the substrate's deformable layers. It is noted that there is a practical limitation on choosing the value of *h* as excessively thick substrate layers can significantly reduce image contrast and resolution due to light scattering (*46*) and refraction (*47*).

In this study, as the noise was up to 1.2 µm in our measurement system after applying a filtering approach (See Supplementary Materials S.6), the PDMS with Young's modulus of 1.4 – 1.7 MPa (*48–50*) and Poisson's ratio of 0.45 was selected to fabricate the substrate at total thickness 300 µm (Figure 2a). For a typical contact stress greater than 0.05 MPa applied on the top surface of the substrate, the ratio of noise to measured deformation $\psi$ is less than 10%.

### 2.3 Validation: A steel ball supported on a PDMS substrate

To evaluate the accuracy of the proposed method, the 3D displacements of the PDMS substrate caused by the self-weight of a 20.6 g solid steel ball that rested on the top of the PDMS (see Figure 4a), were simulated and compared with experimental measurements. The steel ball was selected because it exerted a gravitational force of 0.202 N, which is comparable to the maximum tensile load exerted by marine mussel plaques, discussed in the next Section. This provides a realistic case study aligned with the intended application. To assess the method's performance in a wet environment, the same setup was tested with the ball fully immersed in water. In this case, the net force was reduced to 0.176 N due to an upward buoyant force of approximately 0.026 N.

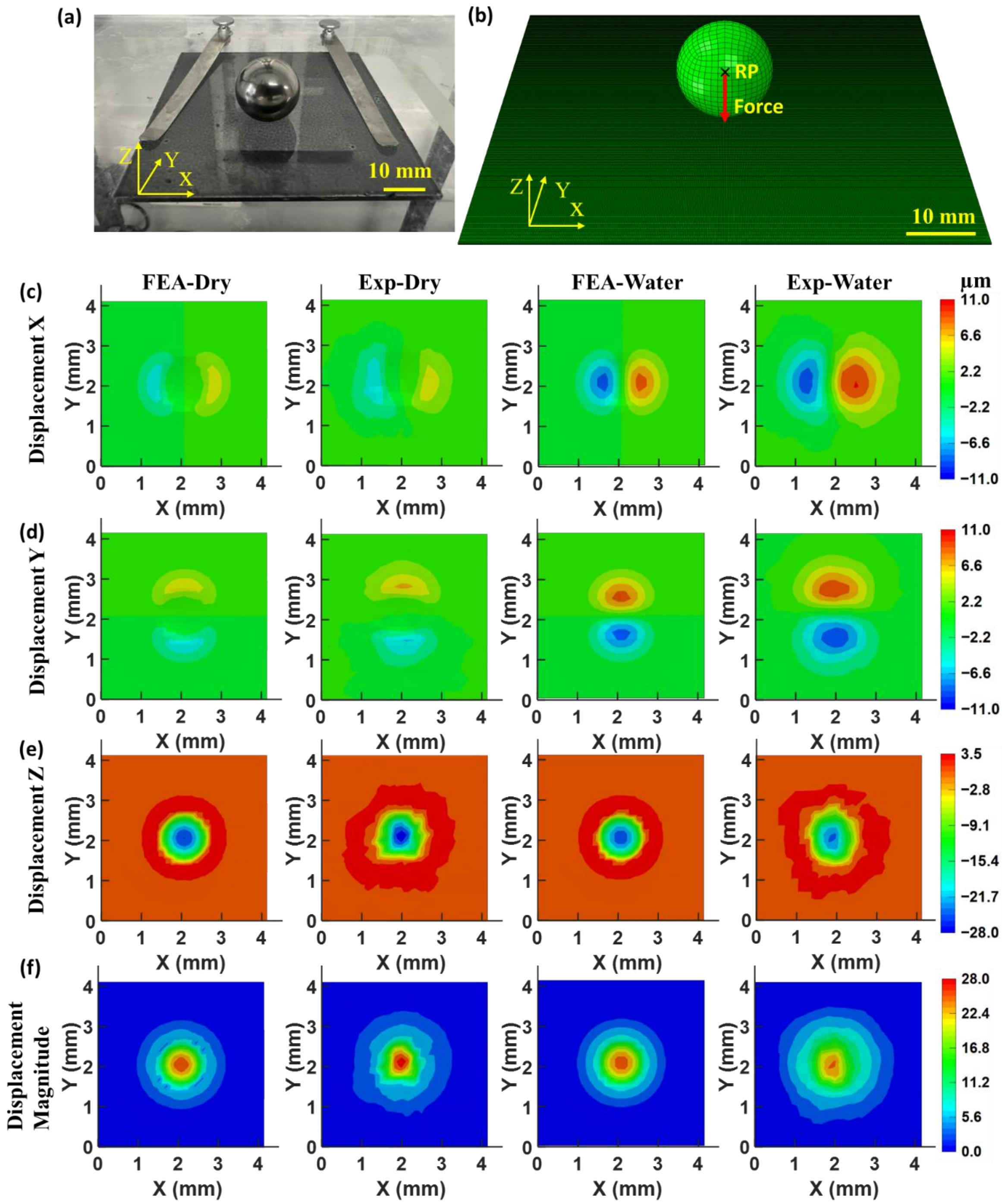

Figure 4. A PDMS substrate under the self-weight of a 20.6 g steel ball: (a) the experimental setup, (b) the FE model, (c) the FE predictions and the SDIC measurements of the displacements in (c) X direction, (d) Y direction, (e) Z direction, as well as (f) displacement magnitude.

Finite element simulations were performed using ABAQUS®/Standard. The steel ball, with a diameter of 17 mm, and the PDMS substrate, with dimensions of 80 × 80 × 0.3 mm$^3$, were modeled to reflect the experimental setup, as shown in Figure 4b. The PDMS substrate was discretized using 8-node brick elements with reduced integration and hybrid formulation, i.e., C3D8RH element in ABAQUS notation. The elements are suitable to model incompressible or nearly incompressible solids. The steel ball was discretized using 8-node brick elements with reduced integration (C3D8R) and modelled as a rigid body by applying constraint equations. A concentrated force was applied at the reference point (RP), located at the center of the ball, in the negative direction along Z-axis, to simulate the indentation load applied to the substrate in the experiment. The bottom surface of the PDMS substrate was fully fixed to replicate the experimental boundary conditions. The contact between the ball and the substrate under dry condition was defined using a surface-to-surface interaction with tangential behavior governed by the penalty formulation, applying a friction coefficient of 1.7 (*51*). Normal behavior was modelled using hard contact. The interaction between the ball and the PDMS substrate under water was modelled using a general contact formulation with a frictionless condition, as the ball–PDMS interface immersed in water experiences negligible friction due to the lubricating effect of the interfacial water layer, which prevents direct solid–solid contact.The effect of friction between the ball and the substrate is discussed in Supplementary Materials S.5. For simplicity, the cohesion between the ball and the substrate was ignored.

The deformations on the top surface of the PDMS substrate are shown in Figure 4 (c-f), along with comparison between FE predictions and experimental measurement. The labels ‘FEA-Dry’ and ‘FEA-Water’ refer to FE predictions under dry and aqueous conditions, respectively, while ‘Exp-Dry’ and ‘Exp-Water’ correspond to experimental measurement under the same conditions. Both the simulated and measured results exhibited a radially symmetric displacement field, attributed to the spherical contact geometry. Initial deformation began at the point of contact and propagated radially with increasing indentation depth, reaching a maximum deformation at the centre. The diameters of the circular deformation patterns remained consistent across both dry and wet conditions. The measured diameter under dry condition (~3.8 mm) were approximately 5% smaller that obtained by the FE prediction (~4.0 mm) (See Supplementary Materials S.8). This discrepancy may be attributed to (1) the characteristics of SDIC measurement, which tracks the average displacement of subsets consisting of multiple particles. As a result, the deformation boundary captured by SDIC may not be accurately estimated, especially near the transition zones where displacements gradually diminish rather than sharply drop to zero; and (2) the cohesion between the ball and the substrate, which was not modelled in the FE predictions.

The measured displacement fields in the X, Y, and Z directions also agree well with those obtained by FE predictions. All the samples exhibited symmetric displacements in the X and Y directions, with peak values of approximately ±4.8 μm (FEA-Dry) and ±5.1 μm (Exp-Dry) - the negative values refer to displacements along the negative directions of the respective axes. The peak displacements in X and Y direction of FEA-Water and Exp_Water are ±8.2 μm and ±9.0 μm, respectively. The maximum X-Y displacements under the wet condition are relative greater than those under the dry condition, which can be attributed to the reduced friction effect in the presence of water. The peak displacement in the Z direction is almost identical to the overall magnitude of displacement, as the substrate deformation was dominated by the vertical load from ball indentation. The results in FEA_Dry and Exp_Dry show peak displacements of 25.5 μm and 28.0 μm in the Z direction, respectively, with a discrepancy of less than 9.8%. For the wet condition, the results in FEA_Water and Exp_Water show peak displacements of 24.5 μm and 23.7 μm, respectively, with a discrepancy of less than 3.4%.

Figure 5 presents the distribution of the traction forces ( i.e., contact forces ) exerted by the steel ball on the top surface of the PDMS substrate, calculated as the nodal forces on the top surface of the FE model of the substrate based on the displacement fields obtained by the FE predictions (i.e., FEA-Dry and FEA-Water), and the SDIC measurements (i.e., Exp-Dry and Exp-Water), respectively. The traction forces between the steel ball and the PDMS substrate are compressive with the peak compressive force occurring at the center of the contact zone, directly beneath the ball, and decreasing radially toward the periphery. This distribution aligns with the displacement field and reflects the spherical geometry of the indenter. The traction force distributions from FEA-Dry and FEA-Water agree well with those from Exp-Dry and Exp-Water, respectively. The measurements (Figures 5b) indicate that the circular area over which the traction forces are distributed has a diameter of approximately 1.55 mm, which is 14.0% larger than the FE prediction of 1.36 mm (Figures 5a). The difference between experimental and FE prediction under wet condition is 10.5%. The peak traction forces are 7.5 mN for FEA-Dry and 6.9 mN for FEA-water, compared to 7.6 mN for Exp-Dry and 6.6 mN for Exp-Water. The resultant traction (contact) forces at the interface, calculated for the Exp-Dry and Exp-Water conditions, are 0.222 N and 0.168 N, respectively—both within 9.9% of the corresponding applied forces (i.e., 0.202 N and 0.176 N). Since the resultant contact forces are derived from element-level stress data, the discrepancy between the predicted and measured forces is close to that observed for displacements (up to 9.8%). Furthermore, for the Exp-Dry condition, the higher mesh densities of 950 and 1900 elements/mm³ were also employed, with the resultant and peak values of the traction forces remaining consistent at 0.184 N and 7.7 mN, respectively, thereby confirming mesh independence. A comprehensive quantitative and visual comparison between the experimental measurements and FE simulations is presented in Supplementary Material S.8.

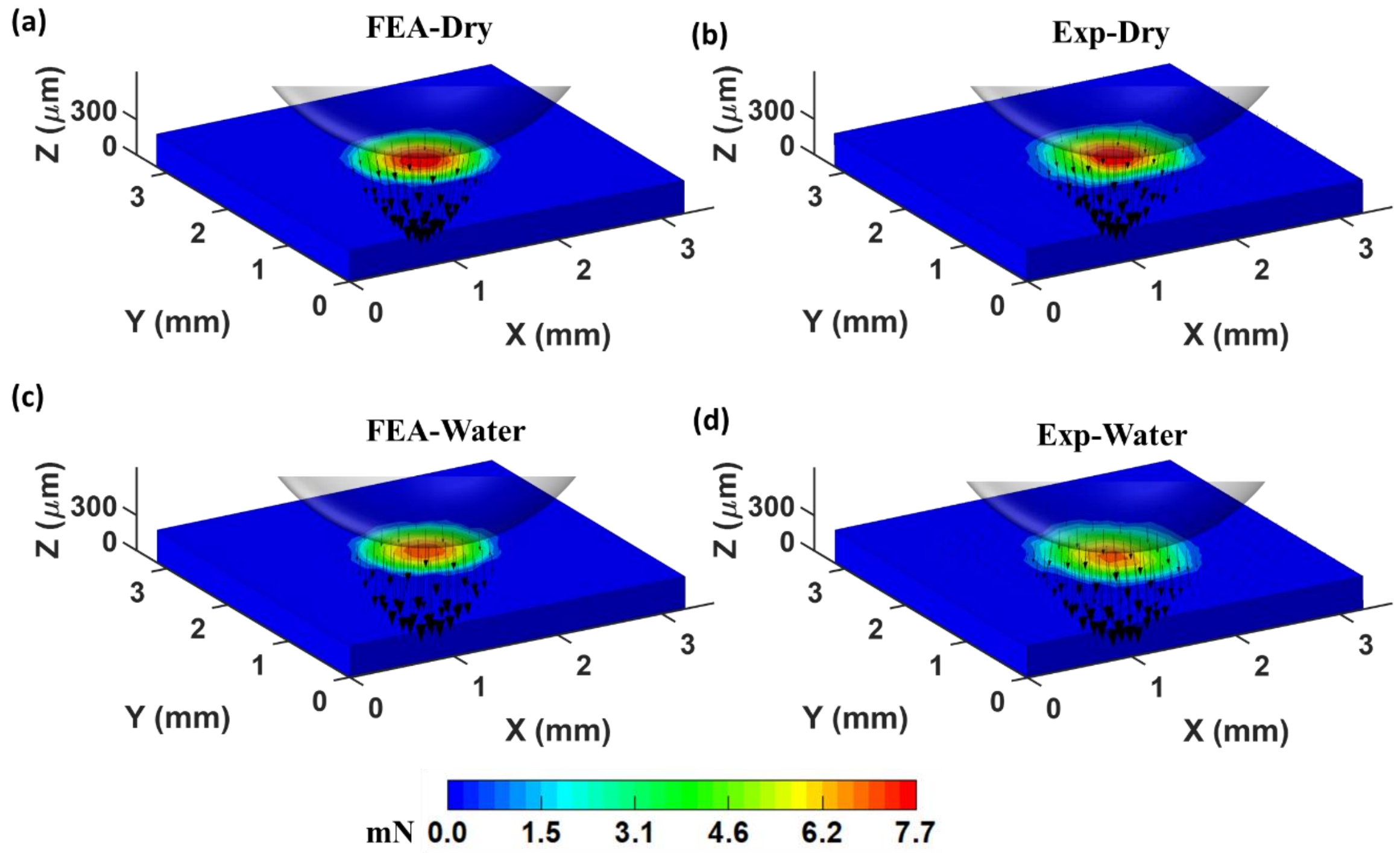

Figure 5. The contact forces exerted by the steel ball on the top surface of the PDMS substrate, calculated based on the displacement fields obtained by the FE predictions, (a) FEA-Dry and (c) FEA-Water; and the SDIC measurements, (b) Exp-Dry and (d) Exp-Water, respectively.

## 2.4 Measurement of the traction forces at wet adhesion between a mussel plaque and a PDMS substrate

Figure 6 illustrates the definitions of key quantities relevant to the test, including the global coordinate system and strains. The global coordinate system (X–Y–Z) is defined such that the X-axis aligns with the projection of the thread onto the substrate surface, the Z-axis is oriented perpendicular to the substrate (normal direction), and the Y-axis is determined accordingly using the right-hand rule. The total strain of the thread–plaque system is defined as the combined elongation of thread ($\Delta L_t$) and plaque ($\Delta L_p$) relative to their initial lengths (thread: $L_t$ = 4.6 mm; plaque: $L_p$ = 267 µm) along the pulling direction, i.e., $(\Delta L_t + \Delta L_p)/(L_t + L_p)$ . The strains of the thread and the plaque are defined as $\Delta L_t/L_t$ and $\Delta L_p/L_p$, respectively.

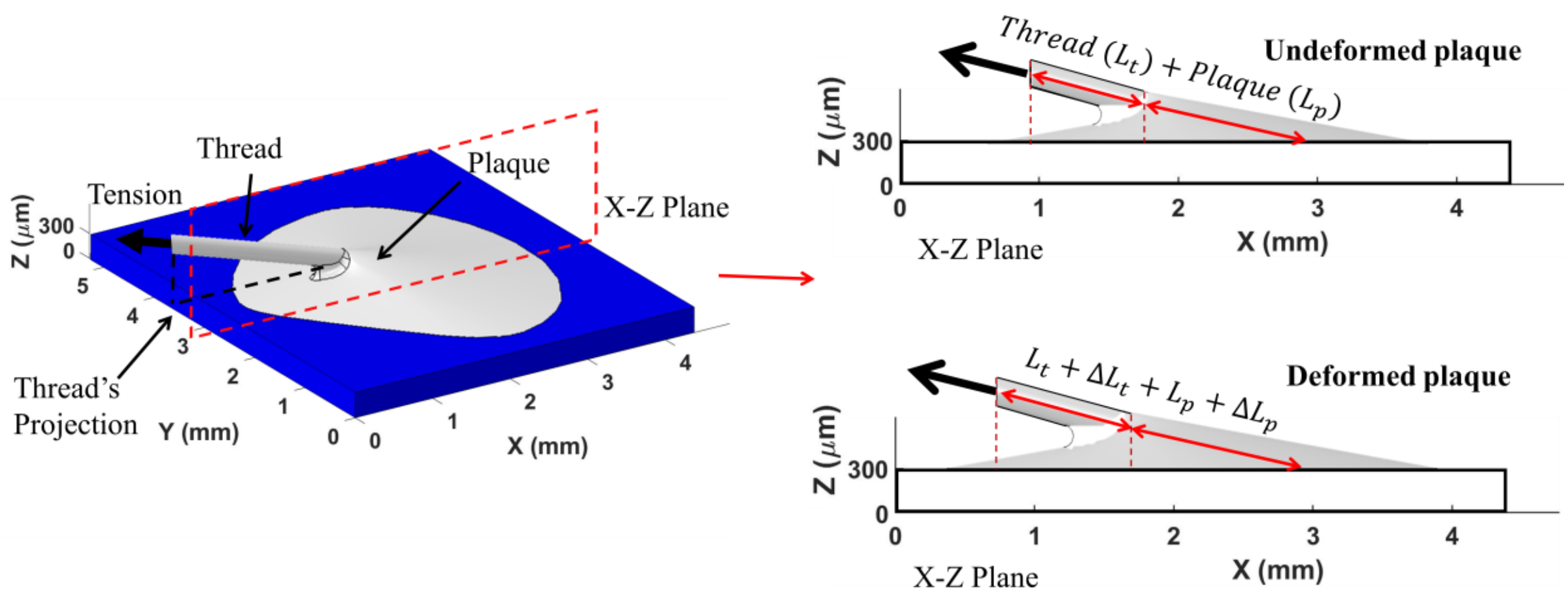

Figure 6. Schematic illustration of the tension test on a mussel thread-plaque system attached to a substrate.

Figure 7a presents the force–extension curves of the mussel thread–plaque system under 15° tension measured by the load cell, including both the raw experimental data (Exp) and corresponding the smoothed curve (Fitted). The curve was smoothed using the Savitzky–Golay filter with a window size of 1501 points and a polynomial order of 3 using the OriginPro software. The resultant traction force (RF) exerted by the mussel plaque on the top surface of the PDMS substrate was calculated as the sum of the reaction forces at the interface. This was determined using the FE model of the substrate based on the displacement fields obtained by the SDIC measurement, shown in Figure 7a for comparison.

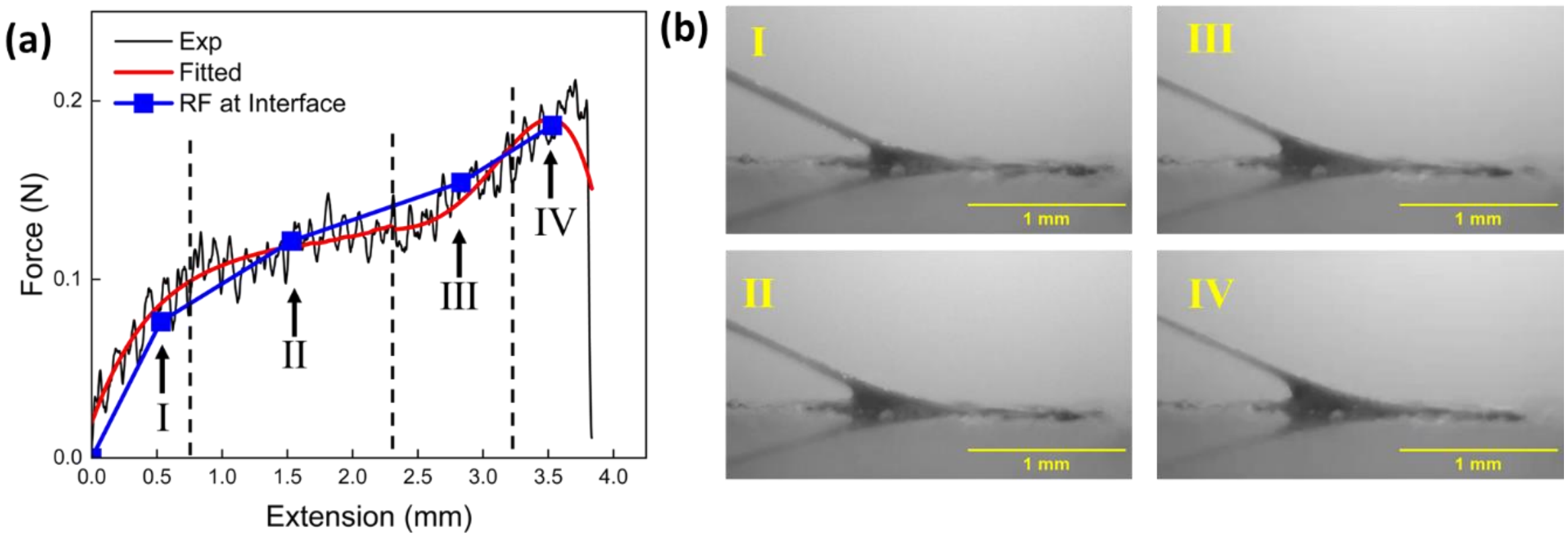

Figure 7. (a) The measured force-extension curve of the mussel thread-plaque system under 15° tension (Exp and Fitted). The resultant traction forces (RFs) at the interface are shown in comparison; (b) The side views of the mussel plaque corresponding to the four stage points.

Four distinct mechanical response stages are identified: linear elastic (Stage I), plateau (Stage II), hardening (Stage III), and failure (Stage IV). In Stage I, the tensile force increases linearly with extension from 0 to 0.8 mm, reaching approximately 0.10 N. Stage II is characterized by a rapid increase in extension (0.8–2.4 mm) with only a slight increase in force (from 0.10 to 0.12 N), indicating a plateau response. This is followed by Stage III, where the system exhibits strain hardening: the force rises to 0.18 N as the extension continues to increase up to 3.3 mm. In the final stage (Stage IV), the tensile force reaches a peak of 0.20 N at an extension of 3.5 mm, followed by a sudden drop, indicating catastrophic failure due to plaque detachment from the substrate. The tensile response of the mussel thread–plaque system reveals complex, multi-stage behavior reflecting intricate interactions within the plaque and at its interface. The extended deformation (see Figure 7b) before failure suggests adaptive interfacial mechanisms that delay detachment. To better understand these effects, exploring the interfacial traction forces at each stage of deformation is essential.

Four representative points were selected along the stress–strain curve: the Stage Point I captures the linear elastic response; the Stage Point II marks the onset of inelastic deformation; the Stage Point III reflects the strain hardening behavior; and the Stage Point IV corresponds to the initiation of failure or material softening. The resulted traction forces of Stage Points I–IV were approximately 0.08 N, 0.12 N, 0.15 N, and 0.19 N, respectively. These values closely align with the fitted force–extension curve of the mussel's tension test, indicating efficient force transmission from the thread–plaque system to the plaque–substrate interface. It is noted that the tensile force measured by the load cell is not necessarily equal to the total traction force imposed on the top surface of the PDMS substrate, as there might be suction force caused by volume change of the plaque (*52*).

The components of resultant traction forces in the X, Y, and Z directions (RF_X, RF_Y, and RF_Z) are summarized in Table 2. The negative sign of RF_X indicates that the force acts along the negative X-axis. The RF_Y values remained close to zero, indicating negligible reaction forces in the Y direction. This observation is consistent with the loading direction, which projected predominantly along the X-axis. The values of RF_X are significantly higher than both RF_Y and RF_Z, suggesting that shear interfacial stress along the X-axis is the dominant mode of interfacial loading. The directions of the resultant reaction force vectors were calculated to range between 16° and 23°, demonstrating good alignment with the applied tensile load direction of 15°. The minor discrepancy is likely due to slight deviations in the experimental setup, particularly in maintaining precise control over the loading angle.

Table 3. The components (RF_X, RF_Y, and RF_Z), magnitudes (RF) and directions of the resultant traction forces at the four stage points.

| Stage Point | RF_X (N) | RF_Y (N) | RF_Z (N) | RF (N) | Direction |
|---|---|---|---|---|---|
| 1 | -0.070 | 0.006 | 0.030 | 0.076 | 23° |
| 2 | -0.116 | 0.008 | 0.035 | 0.121 | 17° |
| 3 | -0.145 | 0.009 | 0.052 | 0.154 | 20° |
| 4 | -0.179 | 0.012 | 0.050 | 0.186 | 16° |

The SDIC measured 3D displacement distribution at the plaque–substrate interface for each stage is shown in Figure 8 (a-d). The high-magnitude displacement was primarily concentrated in the left-front region of the plaque, aligning with the projected area of the thread on the substrate. The maximum displacements for Stage Points I–IV are approximately 18 µm, 30 µm, 38 µm, and 47 µm, respectively. Across all loading stages, the displacement component in the Y-direction displays a symmetric pattern. The X-direction displacement dominates the deformation behaviour, while displacement in the Z-direction remains minimal—limited to approximately 3 µm. The directions of the overall displacements, indicated by black arrows on the substrate, are predominantly along the negative X-axis and slightly lower than the applied 15° tensile direction. This alignment suggests that shear deformation dominates at the interface.

Consequently, the plaque's deformation under 15° tension is governed mainly by interfacial shear rather than normal separation.

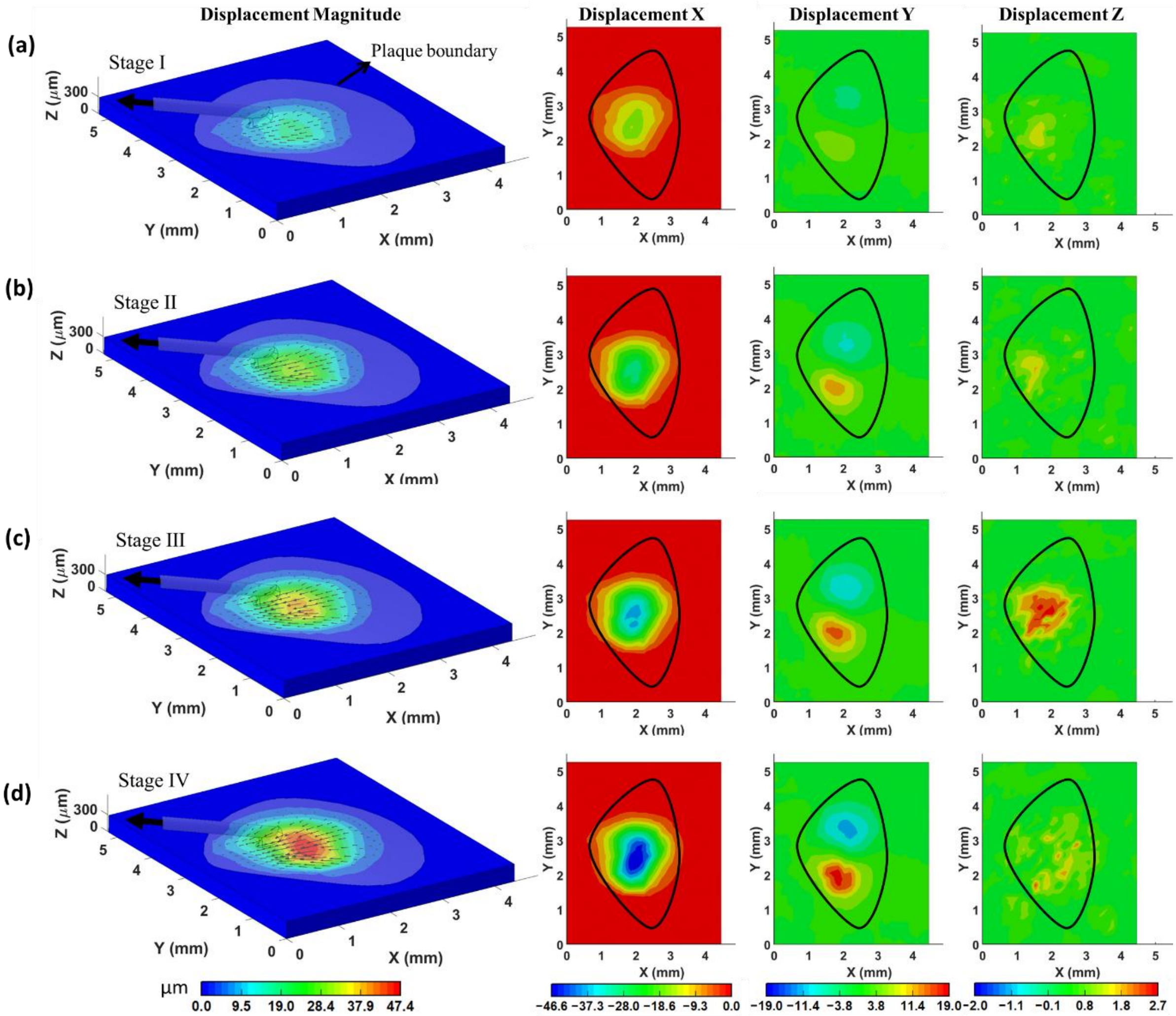

Figure 8. (a-d) The displacements of the substrate at the interface at the four Stage Points of 15° tension.

Based on SDIC measured displacement field on the top surface of the PDMS substrate, the distributions of nodal forces on the top surface, exerted by the mussel plaque through the interface, are shown in Figure 9 (a-d) for both 3D view and sectional views at the four Stages Points. Traction force vectors are shown as arrows originating from interface nodes and pointing outward, indicating the directions of the traction forces. The arrow colours represent the force magnitudes in the 3D space, following a continuous colour scale. A cross-sectional view in the X–Z plane is taken along the central axis of the thread.

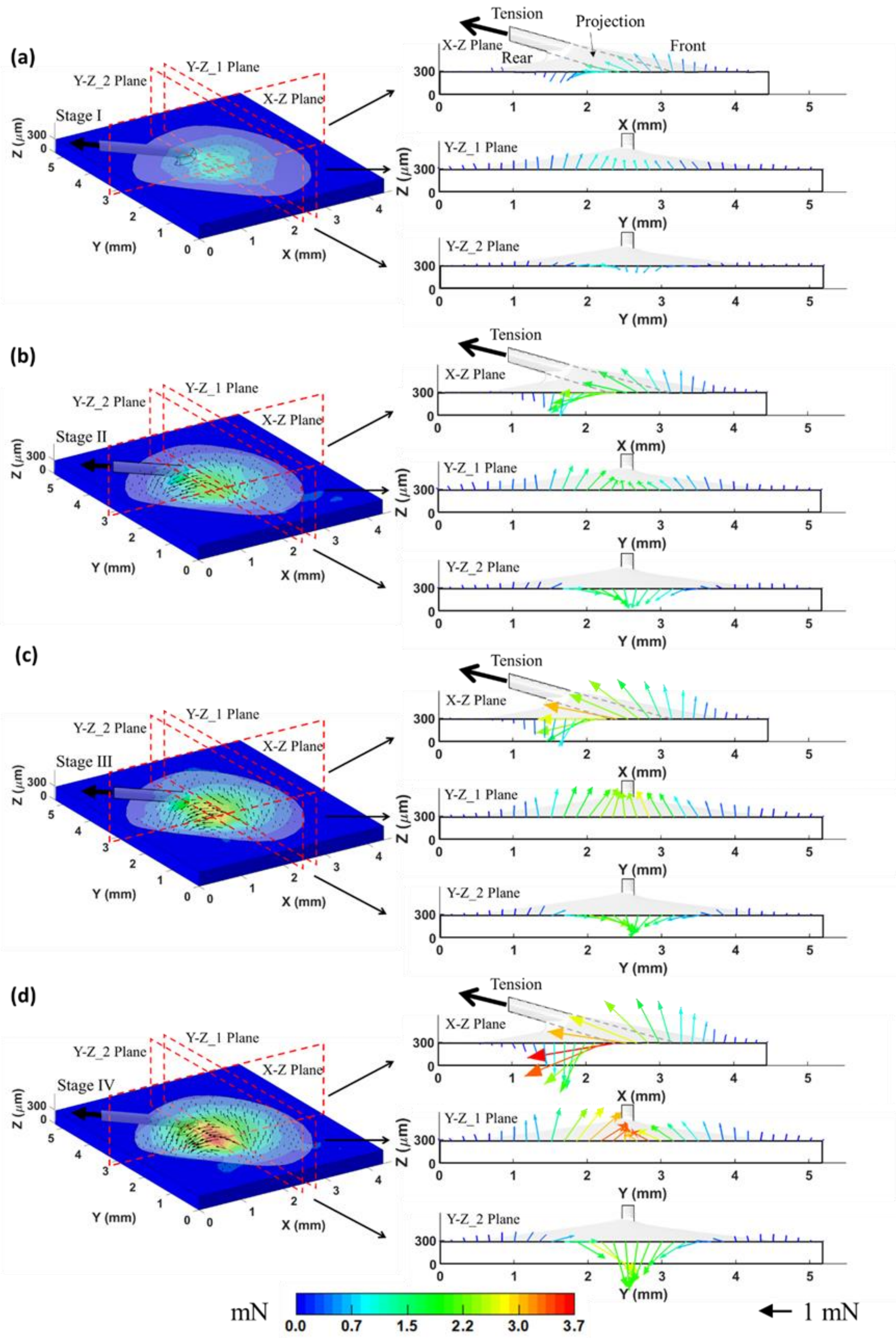

Figure 9. (a-d) The nodal forces on the top surface of the substrate at the four Stage Points of the 15° tensile test.

As shown in the 3D and X–Z cross-sectional views, the distribution of traction forces across the substrate surface demonstrates a clear transition from tensile loading in the anterior region (X > 3 mm) to compressive loading in the posterior region (X < 2 mm), with an intermediate transition zone corresponding to the thread-projected area (2 mm ≤ X ≤ 3 mm). To elucidate the spatial variation of traction forces within this transition region, two Y–Z plane cross sections—positioned near the front (Y–Z_1) and rear (Y–Z_2) of the plaque—are presented in Figure 9. The data indicate that the transition zone extends approximately across 1.6 mm ≤ Y ≤ 3.5 mm. In these sections, the arrow lengths denote the magnitudes of the force components projected onto the respective planes. The analysis reveals that the horizontal component of the applied tensile load is predominantly transmitted to the substrate through the transition zone, whereas the vertical component is primarily transmitted via the anterior region of the interface. Furthermore, the tensile loading induces a bending moment on the interface, which contributes to the simultaneous occurrence of tensile and compressive traction forces within distinct regions of the substrate. These observations highlight the complex nature of force transmission at the plaque–substrate interface, which is crucial for understanding interfacial stability. The biological conclusions regarding mussel adhesion presented here are preliminary, as the mussel plaque adhesion test was intended primarily to demonstrate the applicability of the proposed measurement framework. The ongoing research work (*52*) focuses on involve systematic testing with n ≥ 5–10 biological replicates to ensure statistical robustness and to further elucidate the mechanisms of plaque adhesion under different loading modes.

### 2.5 Effects of the material properties of the PDMS substrate on measurement accuracy

This section investigates the sensitivity of SDIC measurements to the mechanical properties of the substrate. Specifically, the effects of Young's modulus (E) and Poisson's ratio (ν), within their experimentally determined ranges, are systematically evaluated. While substrates in many prior SDIC studies have been modelled as linear elastic solids for traction force computation at the cell–substrate interface, this study adopts the Ogden hyperelastic model to more accurately capture the nonlinear mechanical behaviors of the substrate. This consideration is particularly relevant in the context of mussel plaque–substrate interactions, where elevated traction forces may induce substantial local deformation modes. The influence of the substrate's constitutive model on the accuracy of traction force estimation is critically examined in the analysis that follows.

*Effect of Young's Modulus*

Table 3 summarises the upper bound, average, and lower bound of Young's modulus obtained from data fitting of the experimental tensile data across five PDMS samples produced in a single batch (Supplementary Materials S.2). These correspond to Young's modulus values of 1.43 MPa, 1.38 MPa, and 1.25 MPa, respectively. It is noted that the mean value (1.38 MPa) was employed in the studies reported in Sections 3 and 4.

Table 4. The Ogden models' constants, and corresponding shear and Young's modulus.

| Model | $\mu_1$(MPa) | $\mu_2$(MPa) | Shear modulus (MPa) | Poisson's ratio | Young's modulus (MPa) |
|---|---|---|---|---|---|
| Upper limit | 4.94E-04 | 0.494 | 0.494 | | 1.43 |
| Average | 4.71E-04 | 0.475 | 0.475 | 0.45 | 1.38 |
| Lower limit | 3.89E-04 | 0.432 | 0.432 | | 1.25 |

Sensitivity study has been conducted for experimental studies reported in Sections 3 and 4. Figure 10a shows the deviations between the SDIC measured resultant traction forces (RFs) on the top surface of the PDMS at the interface, using these Young's modulus values, and the applied force (AF) from the self-weight of the steel ball studied in Section 3, i.e., 0.202 N for dry condition and 0.176 N for wet condition. The deviation is calculated as $|RF - AF|/AF \times 100\%$. The findings indicate a moderate sensitivity of the traction force measurements to the substrate stiffness, with deviation values ranging from approximately 1% to 23%. Figure 10c shows the comparison between the SDIC measured RFs and the measured tensile forces (Exp) applied on the thread across all four loading stages as reported in Section 4. The SDIC results were not highly sensitive to variations in Young's modulus: the curves generated using the upper, average, and lower bound modulus values closely matched the experimental tensile data, indicating the robustness of the SDIC method in capturing interfacial force response across a realistic range of material stiffness.

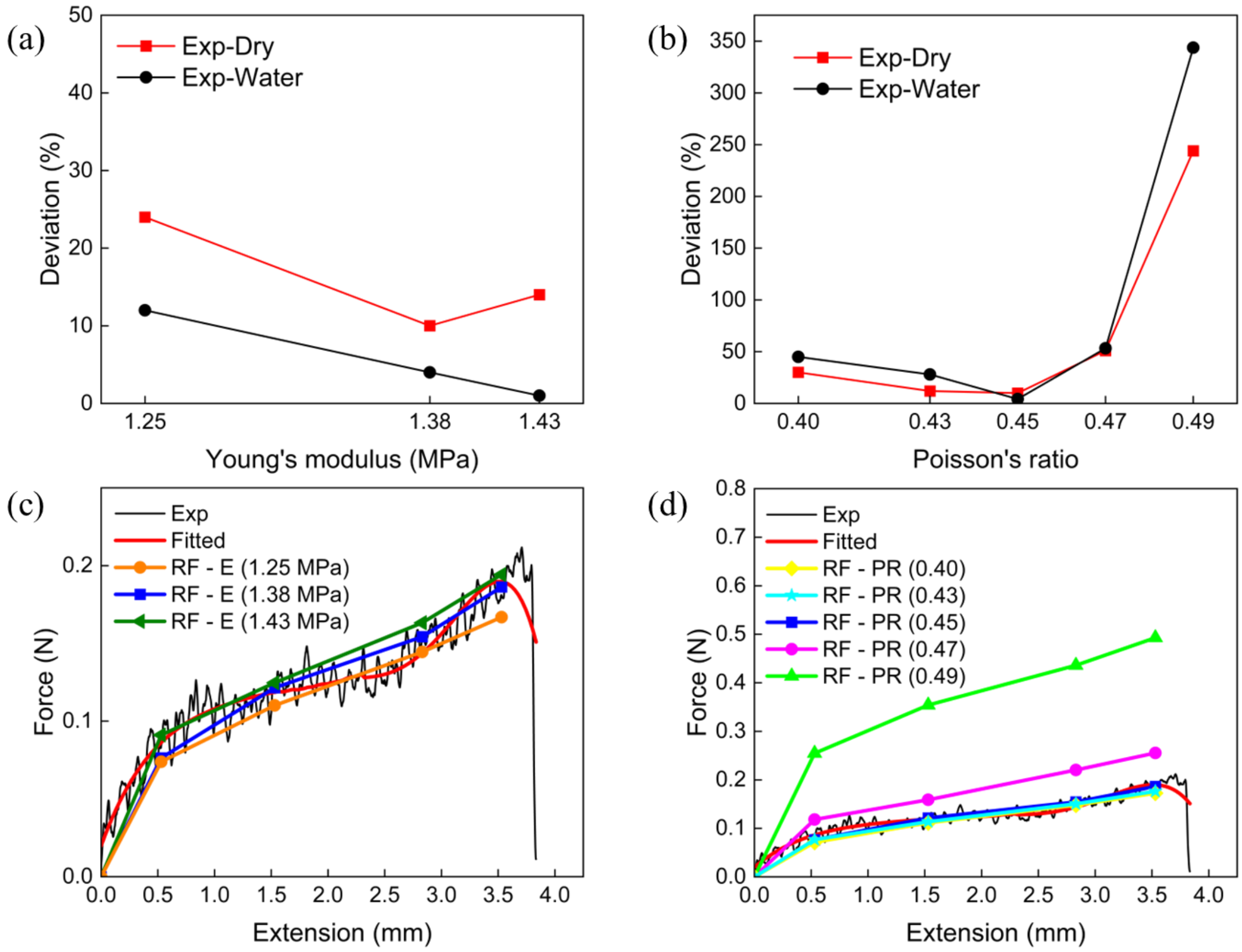

Figure 10. Sensitivity analysis of the PDMS substrate material parameters: (a) and (b) the effects of Young's modulus ( E ) and Poisson's ratio (PR) to the deviation between the SDIC-measured resultant traction forces (RFs) and the applied force (AF) due to the self-weight of the steel ball (Section 3); (c) and (d) the effects of Young's modulus and Poisson's ratio to the SDIC-measured RF in the mussel thread–plaque system under 15° tension (Section 4 ).

*Effect of Poisson's ratios*

As a hyperelastic material, PDMS poses challenges in accurately determining its Poisson's ratio, which can vary depending on factors such as the curing process (*53*), porosity (*54*), strain rate (*55*), and sample thickness (*56*). Consequently, a definitive value is not consistently reported in the literature, with typical values ranging from 0.40 to 0.49 (*41–43*). Figures 10 (b) and (d) present the effect of Poisson's ratio on the SDIC measurements in the current study. In both cases, the resultant traction forces at the interfaces were calculated using Poisson's ratio $\nu = 0.40$, 0.43, 0.45, 0.47, and 0.49, respectively, while maintaining Young's modulus $E = 1.38$ MPa. Poisson's ratio was found to have a substantial impact on the accuracy of SDIC measurements. For Poisson's ratio $\nu = 0.45$, the deviations in resultant traction forces under dry and wet conditions are 9.9% and 4.3%, respectively, relative to the applied force from the self-weight of the steel ball (Figure 10 (b)). The deviations can increase up to 50% or up to 343.7% when the value of Poisson's ratio decreases to 0.40 or increases to 0.49. For the mussel plaque-substrate interaction measurement, the resultant traction forces calculated using $\nu = 0.45$ showed the closest agreement with the tensile force applied on the thread, with deviations of less than 10% across all the four tension stages. In contrast, the deviations increase to up to 52% or 201% if Poisson's ratios $\nu = 0.47$ or $\nu = 0.49$ are employed in the SDIC calculations. Furthermore, it was found that the direction of the resultant traction force was also significantly influenced by the choice of Poisson's ratio: at $\nu = 0.45$, the angle was calculated as 19° ± 4°, closely aligned with the applied tensile direction of 15°; in contrast, the angles were 46° ± 3° and 69° ± 1° at $\nu = 0.47$ and $\nu = 0.49$, respectively. These sensitivity studies indicate that Poisson's ratio of 0.45 is the most suitable choice for the current study.

The sensitivity analysis of material parameters presented in this section underscores the critical importance of implementing a robust calibration procedure to accurately characterise the mechanical properties of the substrate prior to SDIC measurements. Such calibration is essential to minimise errors and ensure the reliability of the experimental results.

Effect of the constitutive model

The constitutive behavior of the PDMS substrate was modelled using the second-order Ogden hyperelastic model, as described in Eq. (1). To evaluate the influence of the constitutive model on the measurement of reaction forces, the substrate was also modelled using a linear elastic model. The elastic modulus was set to 1.38 MPa, consistent with the value obtained from fitting the second-order Ogden model. A Poisson's ratio of 0.45 was adopted, as it provides the most accurate results, as discussed previously. The finite element mesh consisted of reduced integration 8-node solid elements (C3D8R in ABAQUS notation), with mesh density kept consistent with that used in the hyperelastic model described in Section 2.1

Based on the measured displacement fields of the mussel plaque's tension of Stage Points I–IV, the traction forces calculated using the two constitutive models are presented in Figure 11 (a-d). The results show that the traction forces calculated via the linear elastic model were consistently lower than those obtained from the Ogden hyperelastic model across the entire interface, with the peak values from the linear elastic model being 8 – 12% lower than those from the hyperelastic model across all the stage points. Figure 11e presents the relative difference between the resultant traction forces calculated using using linear elastic model ($RF_e$), and the hyperelastic model ($RF_h$), defined as $(RF_e - RF_h)/RF_h \times 100\%$, as a function of the peak values of the maximum principal strain of the substrate at the four Stage Points. The results show that the difference grows in magnitude from around -1% to -5% as strain increases, indicating that the linear elastic model may increasingly underestimate interfacial traction forces as the substrate experiences greater deformation. This trend is consistent with the hyperelastic behaviour of PDMS, which exhibits non-linear elastic deformation. As strain increases, the effective stiffness of PDMS increases, a characteristic captured by the hyperelastic model but neglected by the linear elastic approximation, leading to greater underestimation of interfacial traction forces in high-strain conditions when using the elastic model.

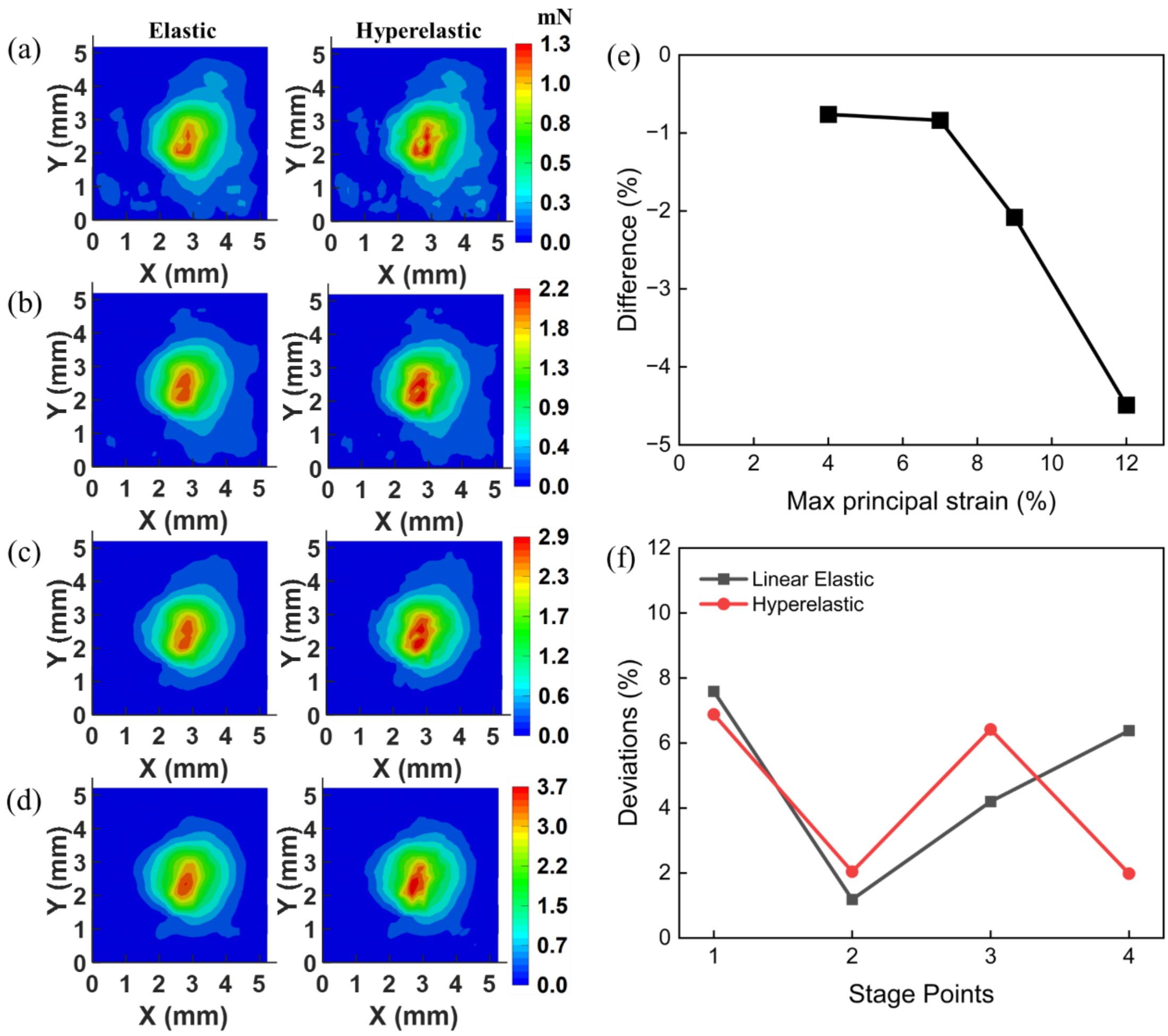

Figure 11. The comparison of traction force distribution calculated using linear elastic and hyperelastic models at Stage Points (a) I, (b) II, (c) III and (d) IV; (e) The difference between the resultant traction forces (RFs) calculated via linear elastic and hyperelastic models as a function of the peak values of max principal strain; (f) the deviations between the resultant traction forces (RFs) and the applied tension forces at the four stage points.

Although the linear elastic model underestimates the interfacial traction forces compared to the hyperelastic model, the deviations relative to the applied tension forces remain below 8%, comparable to those obtained using the hyperelastic model, as shown in Figure 11f. The similarity in deviations suggests that both models may be suitable for small-strain conditions, where the maximum principal strain does not exceed 12%.

However, a critical distinction arises when examining the spatial distribution of these forces. To investigate this, the local traction force magnitudes calculated by the two models, i.e., $F_{hyperelastic}$ and $F_{elastic}$, were compared along the central axes of the mussel plaque interface (Stage IV). Figure 12a plots the force magnitudes derived from both models using the same displacement fields. The plots clearly show that the traction forces of the two models are different. A quantitative calculation of the deviation, defined as $(F_{hyperelastic} - F_{elastic}) / F_{elastic}$, reveals significant discrepancies. Along the central X-axis, the deviation ranged from -8.6% to 37.2%, while along the central Y-axis, it ranged from -6.3% to 15.1%. These significant local fluctuations indicate that while the Linear Elastic model approximates the total force reasonably well, it distorts the stress distribution profile, particularly in regions of complex deformation geometry.

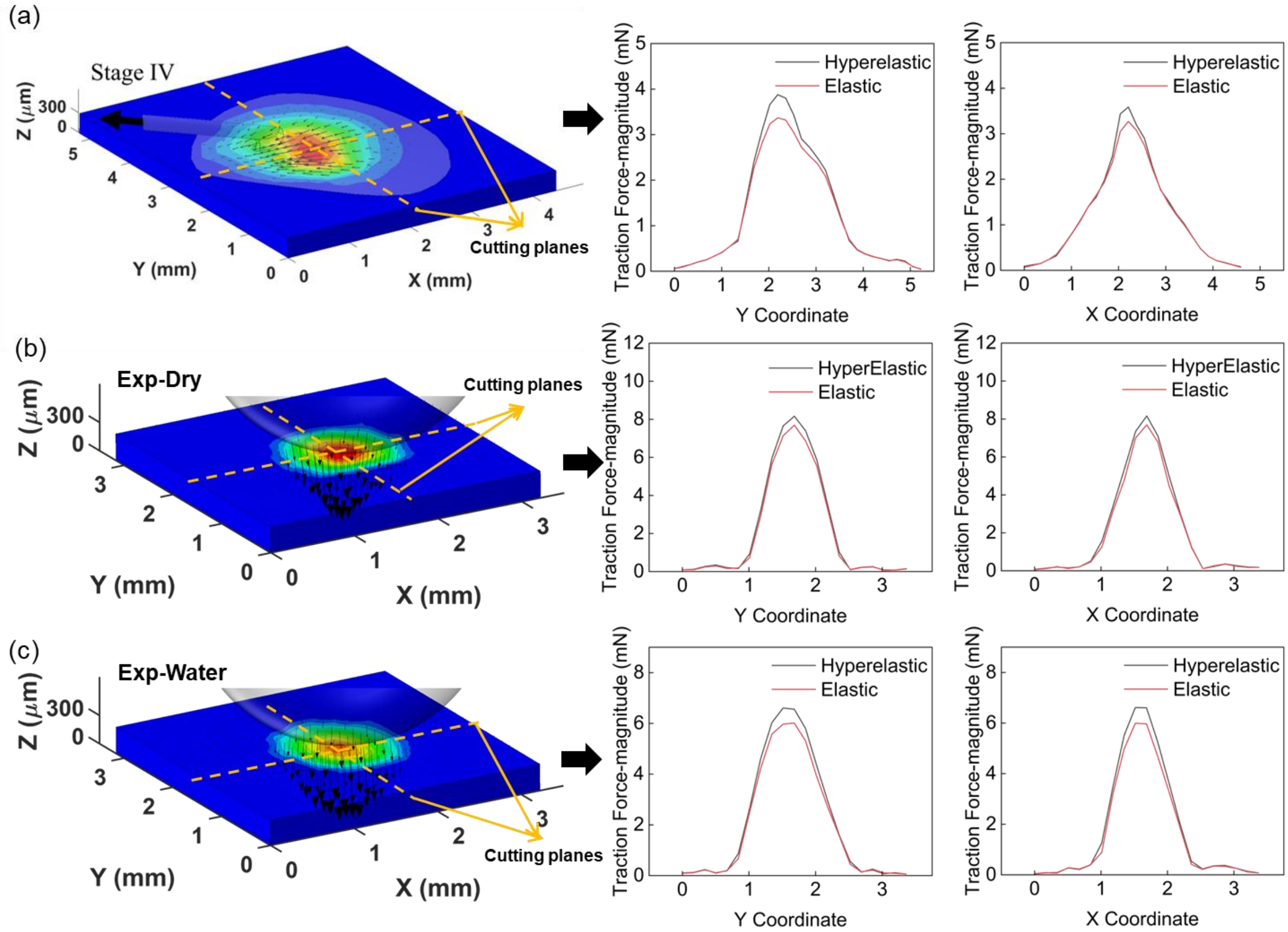

Figure 12. Comparison of traction force magnitudes calculated using hyperelastic versus linear elastic models for (a) mussel plaque tension at Stage IV: Cut views at central axes; Steel Ball experiment under Dry condition: (b) Cut views at central axes (c) Wet condition: Cut views at central axes.

To validate this observation in a geometrically controlled system, a similar comparative analysis was performed using the steel ball indentation data. The identical DIC-measured displacement fields were applied to both the Linear Elastic and Hyperelastic FE models. The resulting traction force magnitudes are plotted in Figures 12b and 12c, visually confirming the discrepancy between the models. For the dry condition (Figure 12b), the local deviation defined as ($F_{hyperelastic}$ - $F_{elastic}$) / $F_{elastic}$, varied significantly across the contact zone. Along the central Y-axis, the deviation calculated from the profiles ranged from -9.0% to 16.9%, while along the central X-axis, it ranged from -20.7% to 25.0%. For the wet condition (Figure 12c), the discrepancies were further amplified. Deviations along the Y-axis ranged from -39.9% to 23.0%, and along the X-axis from -10.3% to 29.5%.

These results, combined with the theoretical validation provided in Supplementary Material S.9, confirm that the linear elastic approximation introduces spatially non-uniform errors that can exceed 30% locally. By failing to account for the stiffening and large-deformation characteristics of PDMS, the linear model compromises the spatial accuracy of the force map. Consequently, the employment of the hyperelastic model is essential for the precise quantification of 3D interfacial traction force distributions in soft bio-adhesive systems.

## 3. Conclusion

This study introduces a robust and versatile traction force microscopy framework for quantifying three-dimensional interfacial traction forces during bio-adhesion events. This technique enables precise 3D quantification of microscale displacements and force distributions at interfaces under both dry and wet conditions. It overcomes the limitations of existing techniques such as atomic force microscopy, fluorescence microscopy, and traction force microscopy, which are often constrained by limited measurement scales and restricted fields of view. Moreover, these methods typically require direct mechanical contact or the embedding of fluorescent beads, both of which can disturb delicate interfaces or influence their inherent mechanical behavior. Key findings and contributions include:

- A theoretical model was developed to determine the optimal thickness of the deformable substrate in the SDIC system, accounting for measurement noise in both normal and shear deformation modes. Based on the model predictions, a PDMS substrate with tailored mechanical properties—specifically, a Young's modulus in the range of 1.4–1.7 MPa and a thickness of 300 μm—was employed in the present study. This configuration was selected to minimise the noise-to-signal ratio (maintained below 10%) while preserving the optical transparency required for accurate displacement tracking.

- The SDIC system was calibrated and validated using a steel ball compression test, which may serve as a standardized procedure for future SDIC applications. In the current study, the experimentally measured 3D displacement fields agreed well with FE simulations, with deviations of less than 6.5% under dry conditions and 3.4% under wet conditions. Furthermore, the resultant traction forces derived from SDIC measurements deviated by 8.8% and 4.3% from the applied force (i.e., the self-weight of the steel ball) in dry and wet environments, respectively.

- The SDIC system successfully quantified the spatial distribution of traction forces at the interface between a marine mussel plaque and a substrate, demonstrating its capability to capture complex interfacial mechanical responses in bio-adhesion events.

- A sensitivity analysis of substrate material parameters has been conducted, which suggests that both Young's modulus and Poisson's ratio have an impact on the SDIC measurement, with Poisson's ratio playing a more pronounced effect. These findings highlight the importance of implementing a robust calibration procedure to ensure the reliability and validity of SDIC measurements.

While many previous SDIC studies have modelled substrates as linear elastic solids for computing traction forces, the present framework incorporates the substrate's hyperelastic behaviour into the traction force calculations. Numerical evaluation showed that (1) traction forces obtained using the linear elastic model were lower than those computed with the Ogden hyperelastic model, and (2) the discrepancy was negligible for small substrate deformations but became more pronounced as deformation increased.

## 4. Experimental Section

### The SDIC measurement

In this study, an experimental platform was developed based on the in-situ SDIC method, shown in Figure 13. A tension device consisting of a linear actuator (Thomson MLA11A05) and a load cell (Honeywell Model 34) was used to apply tension force on an object that was attached to the PDMS substate. Meanwhile, two CCD cameras (DCC1545M-GL, ThorLabs, Exeter, UK) with an angle φ (~30°) were positioned at about 21.5 mm distance relative to the substrate to capture the movement of particles. Both the CCD cameras operated at a frame rate of 25 fps with a resolution of 1280 × 1024 pixels. A side camera (Pixelink PL-D753MU, Edmund Optics, York, UK) was also employed to monitor the deformation of the object in Z-Y plane at the frame rate of 50 fps. The in-situ SDIC system was mounted on a honeycomb optical breadboard (Newport Corporation, California, USA) to control the vibration induced measurement noise to less than 2 μm. A white light source was positioned 11.5 mm below the substrate to enhance the contrast between the particles and the surrounding black-dyed PDMS.

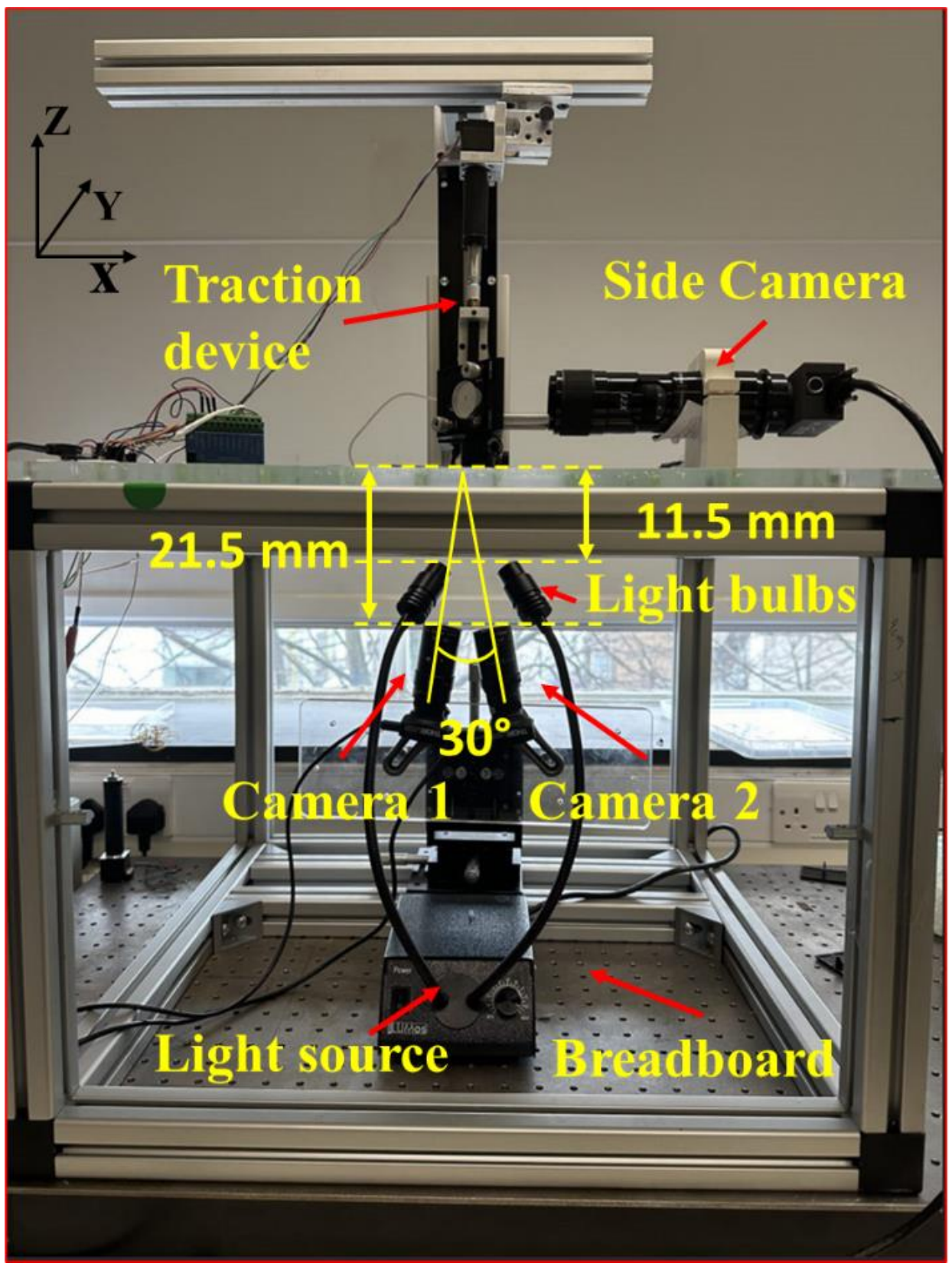

Figure 13. The experimental platform employed in the current study

During the SDIC measurement, the region of interest (ROI), as shown in Figure 14a, was centred in the imaging system to minimize edge aberrations caused by lens distortions. The ROI was divided into identical square subsets, and each subset was distinguishable due to its unique grey-value distribution, which in an 8-bit digital image is scaled from 0 (black) to 255 (white). Each subset contains a unique grey-level distribution, which is essential for the DIC algorithm's tracking and calculation function. The algorithm locates the subset's new position in subsequent images to precisely measure displacement and strain fields across the specimen. Furthermore, the subsets could overlap or remain separate depending on two parameters: subset size ($L_s$) and subset interval ($L_i$). The subset size represents the dimension (in pixels) of a subset within the ROI, while the interval represents the distance between the central points of adjacent subsets.

Using the DICe software, the 3D translations of each subset were tracked through stereo digital image correlation. The X-Y-Z coordinates of these subsets were detected, and deformation was determined by tracking the motions of all subsets, represented by their central points (Figure 14b). The subset size and interval, defined in pixels via DICe, affected measurement accuracy (*57*). $L_s$ should be large enough to include enough speckles for accurate tracking, while $L_i$ determines the resolution of the measurement, avoiding potentially missing localized deformation features. If $L_i$ is smaller than $L_s$, it would result in overlapping subsets (Figure 14c). It is noted that the number of subsets along a length $L_k$ (unit: pixels) was determined solely by $L_i$, i.e., $L_k/L_i$.

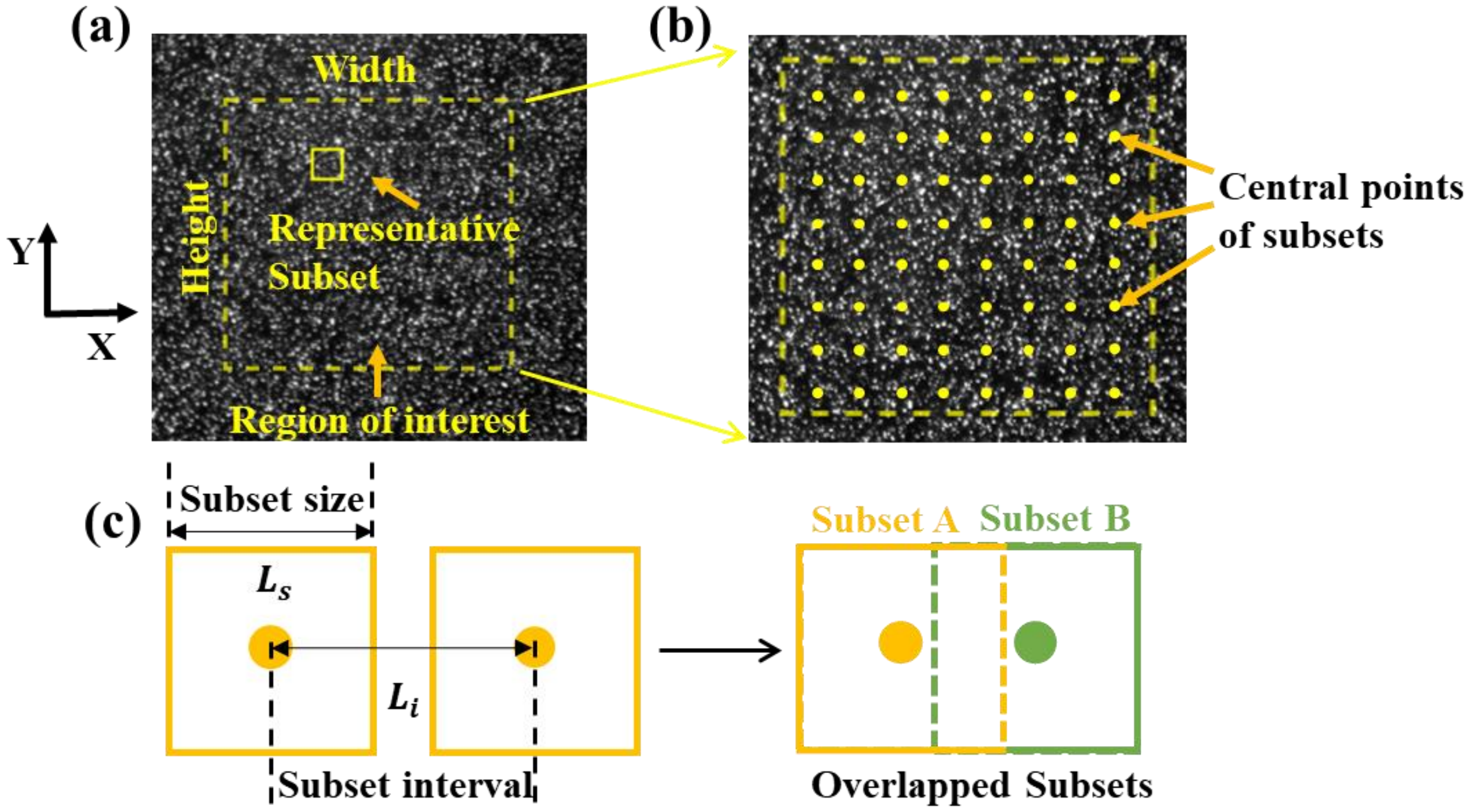

Figure 14. (a) A representative image taken during SDIC measurement; (b) Subsets distribution in the ROI; (c) Schematic showing the definition of $L_s$ and $L_i$.

A parameter study was conducted using DICe, selecting subset sizes ($L_s$=15, 25, 35, 45 pixels) and intervals ($L_i$=10, 15, 25, 35, 45 pixels) from the available options, to evaluate measurement deviations. The parameter study evaluated the deformation of a PDMS substrate under a 9 g steel ball to determine the optimal values of $L_s$ and $L_i$. The results suggest that the deviation of measurement is the lowest when $L_s$ and $L_i$ are 45 and 10 pixels, respectively. The details are summarised in Supplementary Materials S.7. A larger subset size generally improves measurement precision by reducing displacement noise but at the cost of lower spatial resolution. Conversely, a smaller interval increases the density of measurement points and captures finer deformation features, though it also raises computational demands and may increase noise. For balanced performance, it is recommended to select subset sizes of at least 21 pixels to minimize noise, and use intervals smaller than half the subset size, ensuring sufficient overlap without excessive redundancy (*58*). $L_s$ = 45 pixels and $L_i$=10 pixels were used in DIC process in the following studies.

**Tension on the marine mussel plaque**

Blue mussels (*Mytilus edulis*) were collected from the Hunstanton mussel farm (52.94°N, 0.49°E), England. The mussels were secured onto PDMS substrates in an aquarium system with recirculating seawater. Tensile tests were performed on mussel plaques deposited on the PDMS surface within one week of the plaques' deposition. During testing, the plaques remained submerged while their attached threads were pulled at a 15° angle using a linear actuator (MLA11A05, Thomson, Bideford, UK) operating at a quasi-static speed of 0.1 mm/s. The applied load was recorded using a load cell (Honeywell Model 34) with a precision of 0.01 N throughout the tension period. It is noted the natural pulling angles of a mussel thread-plaque system typically occur at around 15°.

The tensile loading angle was set at 15° to simulate a physiological loading condition dominated by shear with a minor normal component, mimicking the hydrodynamic drag forces experienced by mussels in natural environments (*59*). This configuration is also consistent with recent mechanical studies emphasizing the importance of directional tension in determining the quasi-static response and detachment mechanics of mussel plaques (*48*). Consequently, this angle was selected to provide a representative mixed-mode loading scenario to validate the measurement framework.

**Acknowledgment**

The authors thank Dr. Huanming Chen for his assistance with the experimental measurement of the Poisson's ratio of the PDMS substrate, and Dr. Yong Pang for helpful discussions regarding this study.

**Funding:** This work was funded by the Leverhulme Trust Research Grant Scheme, UK (No. RPG-2020-235) and EPSRC Research Grant (EP/X017559/1).

**Author contributions:** Conceptualization: Y.H. and T.L. Data collection: Y.H. Formal analysis: Y.H. and T.L. Funding acquisition: T.L. Investigation: Y.H. and T.L. Methodology: Y.H. and T.L. Simulation: Y.H. and T.L.  Project administration: T.L. Supervision T.L. Validation: Y.H. and T.L. Visualization: Y.H. Writing—original draft: Y.H. Writing—review and editing: Y.H., F.W., and T.L.

**Supplementary Materials**

**S.1 Substrates manufacturing**

The PDMS substrates were manufactured via spin coating process described in Table S1.

Table S1. The spin coating process for manufacturing PDMS substrate

| Step | Materials | Coating speed (rpm) | Coating time (s) | Thickness (μm) |
|---|---|---|---|---|
| 1 | PDMS | 500 | 30 | 125 |
| 2 | PDMS | 500 | 30 | 125 |
| 3 | Particle/PDMS | 2100 | 260 | 15 |
| 4 | Pigment/PDMS | 2100 | 260 | 15 |

The formulation of the monomer and curing agent was 10:1 wt%. The PDMS layers were coated separately on an acrylic substrate via a spin coater (SPIN150i, POLOS, Germany). The coating speed (500 rpm) and coating time (30 s) were chosen for manufacturing the pure PDMS substrate with a thickness of approximately 200 – 300 μm. The third layer was made of PDMS with white micro particles and was spin-coated under 2100 rpm for 260 s. The fourth layer was spin coated with PDMS which was dyed with black silicone pigment (Easy Composites, UK) with 1 wt% under the same coating speed and time as the third layer. The cross-section of the substrate was examined by a microscope (Leica DMI4000B, Leica Microsystems, Wetzler, Germany).

**S.2 The tensile property of the PDMS substrate**

The tension test on the particle-pigmented PDMS was conducted with the tensile strain up to about 25%. The maximum strain in our studies (accuracy validation and case study) was up to 13%. The strain-stress curves between the pure PDMS and the particle-pigmented PDMS are close, and almost identical in the strain range of 0-13%, shown in Figure S1. This indicates that the addition of particles and pigment has negligible effect on the mechanical performance of the PDMS.

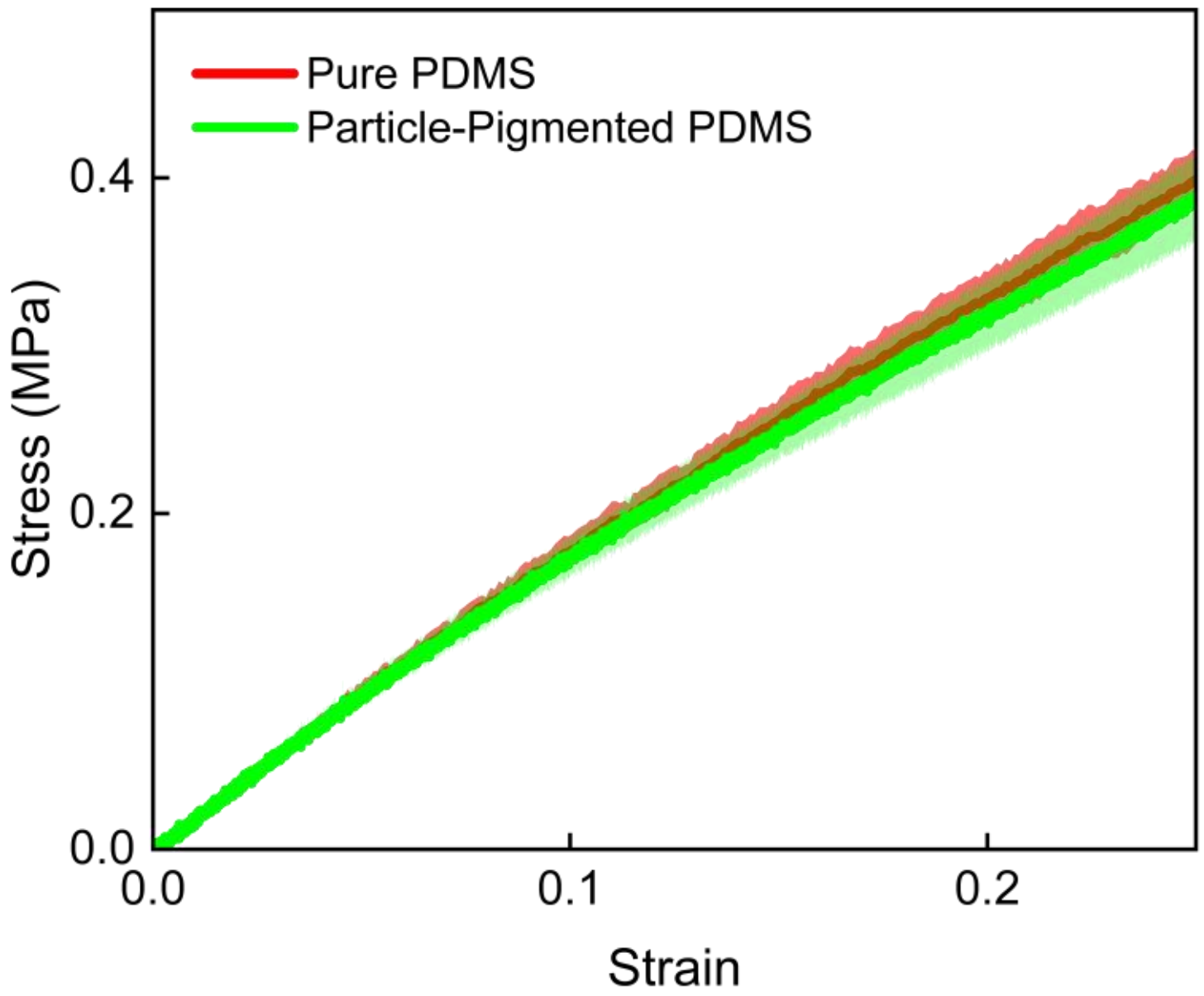

Figure S1. The stress-strain curves of pure PDMS and particle-pigmented PDMS

**S.3 Experimental Determination of the Poisson's Ratio of PDMS Substrate**

To experimentally validate the Poisson's ratio used in the finite-element reconstruction, quasi-static uniaxial tensile tests were conducted on thin PDMS films directly cut from the same speckled substrate used in the traction experiments, including both the ZnS:Cu particle and black pigment layers. The tensile specimens were cut from the PDMS substrate using a hydraulic press (SEALEY, UK) and a precision metal mould conforming to the ASTM D638 Type V geometry.

Quasi-static tests were conducted using a screw-driven load frame (Instron 5980, 1 kN load cell) at 20 °C and a displacement rate of 0.1 mm $s^{-1}$, corresponding to a nominal strain rate of approximately 0.01 $s^{-1}$. Each specimen surface was spray-coated to create a random speckle pattern suitable for DIC analysis, as shown in Figure S2a.

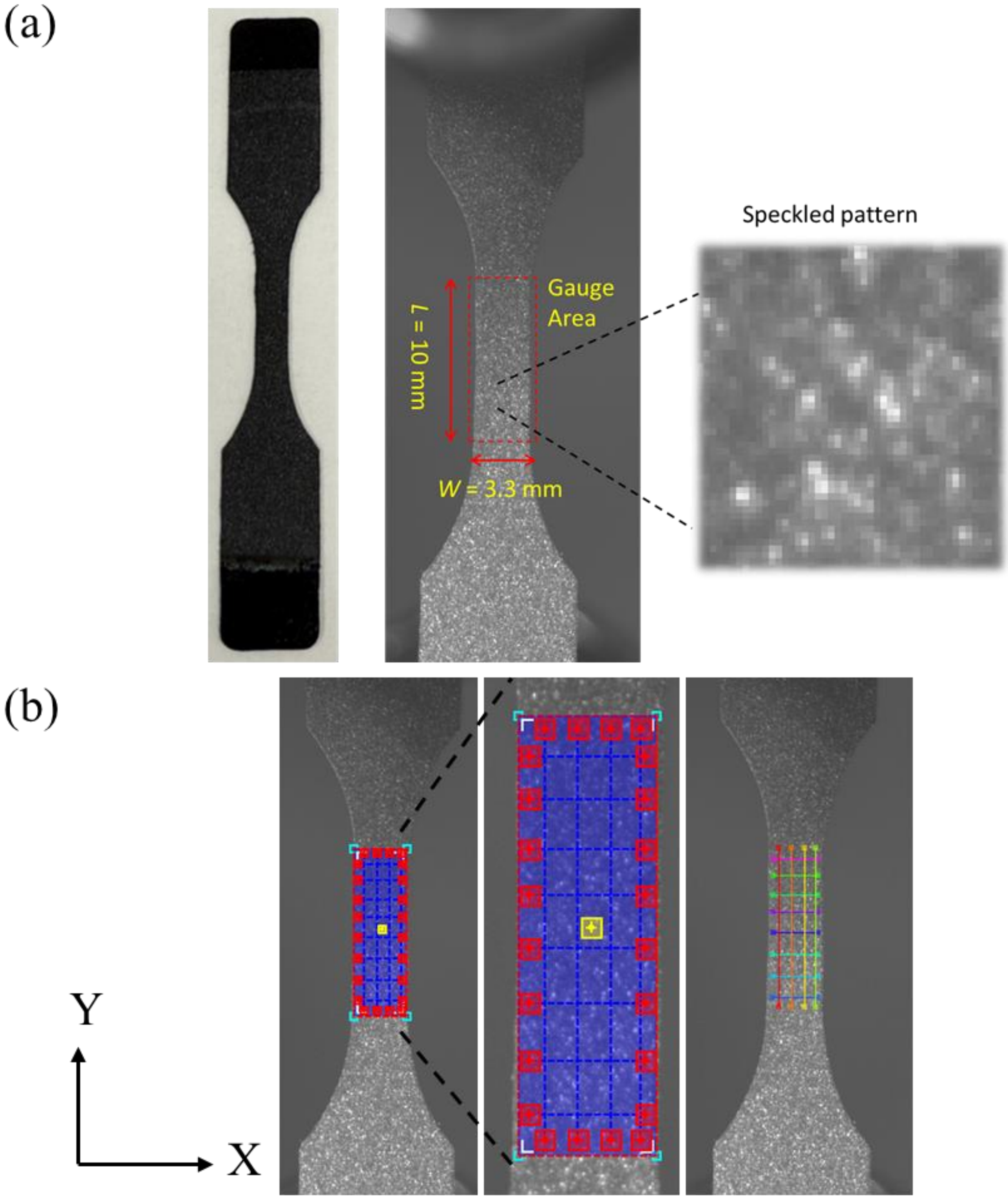

Figure S2. (a) Dog-bone specimen cut from the PDMS substrate for tensile testing, showing the applied random speckle pattern on the specimen surface; (b) Virtual strain gauges positioned along the longitudinal and transverse directions within the gauge section for DIC strain analysis.

Full-field strain fields were recorded using a Point Grey USB 2.0 CCD camera equipped with a Nikon 105 mm lens. The DIC analysis was performed in MatchID (version 2024.1.1) using an affine subset shape function and a zero-normalized sum of squared differences (ZNSSD) correlation criterion. The experimental and DIC parameters are summarized in Table S2. The strain window was set to 113 pixels, which is approximately equal to the width of the gauge section, allowing for a smoother strain field and to minimize noise.

Table S2. The DIC parameters for Poisson’s ratios measurement

| | |
|---|---|
| Image resolution (pixel) | 576 × 2048 |
| Field of view (pixel) | 121 × 1019 |
| Pixel-to-mm conversion | 0.024 |
| Speckle size (pixel) | 5-8 |
| Subset size (pixel) | 17 |
| Step size (pixel) | 4 |
| Strain tensor | Hencky (True) strain tensor |
| Strain window (pixel) | 57 × 57, Quadratic Quadrilateral |

The transverse ($\varepsilon_{transverse}$) and longitudinal ($\varepsilon_{longitudinal}$) true strains were extracted from the gauge section using virtual strain gauge (Figure S2b), and the Poisson’s ratio was computed up to an engineering strain of 1.5 according to

$$v = -\frac{\varepsilon_{transverse}}{\varepsilon_{longitudinal}}$$

The evolution of transverse and longitudinal true strains, together with the corresponding Poisson’s ratio, is shown in Figure S3. The measured Poisson’s ratio within the gauge section, excluding edge regions that may be affected by noise, was ν=0.45 ± 0.03. The selected value of ν=0.45 used in the finite-element reconstruction was further validated by comparing the total reconstructed traction force with the known gravitational force of a reference steel ball described in Section 2.3, demonstrating quantitative agreement within experimental uncertainty. This confirms that the adopted Poisson’s ratio accurately represents the mechanical response of the PDMS substrate used in this study.

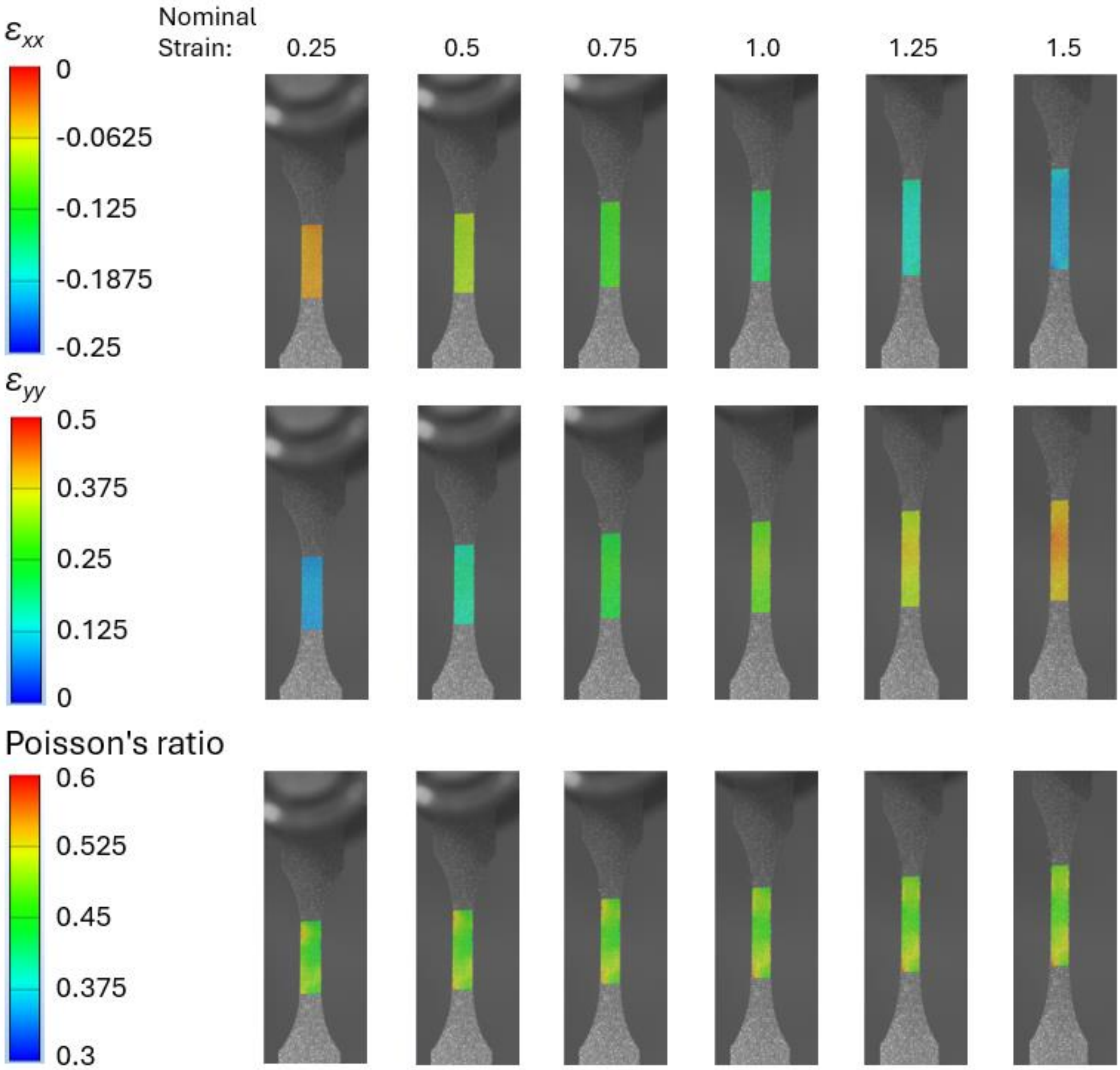

Figure S3. Evolution of true strain in the transverse ($\varepsilon_{xx}$) and longitudinal ($\varepsilon_{yy}$) directions, together with the corresponding Poisson's ratio, as a function of nominal strain up to 1.5.

**S.4 The contact angle and bonding strength of the substrate**

The contact angle was measured using DSA100 (KRUSS, Germany) to investigate the wettability of pristine PDMS and the speckled substrate. Droplets of deionized water were randomly dropped on the surface of the materials and the contact angle between the droplets and the materials were analysed as shown in Figure S4. Five measurements were conducted for each material. The contact angles for pristine PDMS and speckled substrate are 109.2° ± 4.7° and 110.2° ± 6.1°, respectively, showing no significant difference in macroscopic wettability between the two surfaces. This indicates that the addition of speckle particles and black pigment did not substantially alter the surface energy of PDMS.

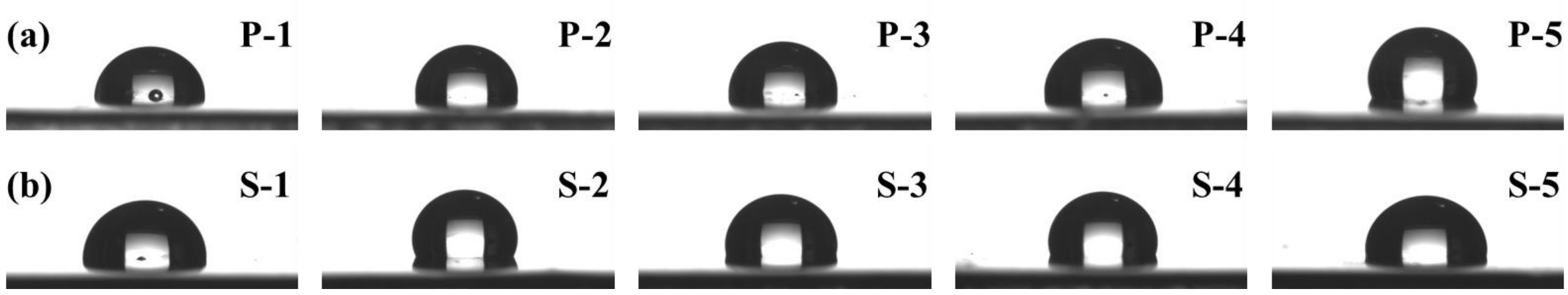

Figure S4. Contact angle images of (a) pristine PDMS samples (P-1 to P-5) and (b) speckled PDMS samples (S-1 to S-5).

To ensure reproducibility and isolate the effect of the surface modification, adhesion performance was compared using a standard 180° peeling test (ASTM D3330) with a commercial bonding tape (3M GPT-020F, RS, UK) rather than mussel plaques, since biological adhesion strength can vary considerably with mussel size, secretion activity, and environmental history (*60*). Tape segments with a total length of 300 ± 1 mm and a width of 25 ± 0.1 mm were bonded to the top surface of the PDMS substrates. As described in Section 2.1, the substrates consisted of either pristine or speckled PDMS mounted on an acrylic panel. The bonded length of each specimen was approximately 65 ± 2 mm. Peeling tests were performed using an Instron 68TM-10 testing machine at a loading rate of 5 mm/s. One grip clamped a folded 12 mm section at one end of the tape, while the other grip held the PDMS substrate. The experimental setup is illustrated in Figure S5a. Each test was repeated three times for both substrate types, and the results are summarised in Figure S5b.

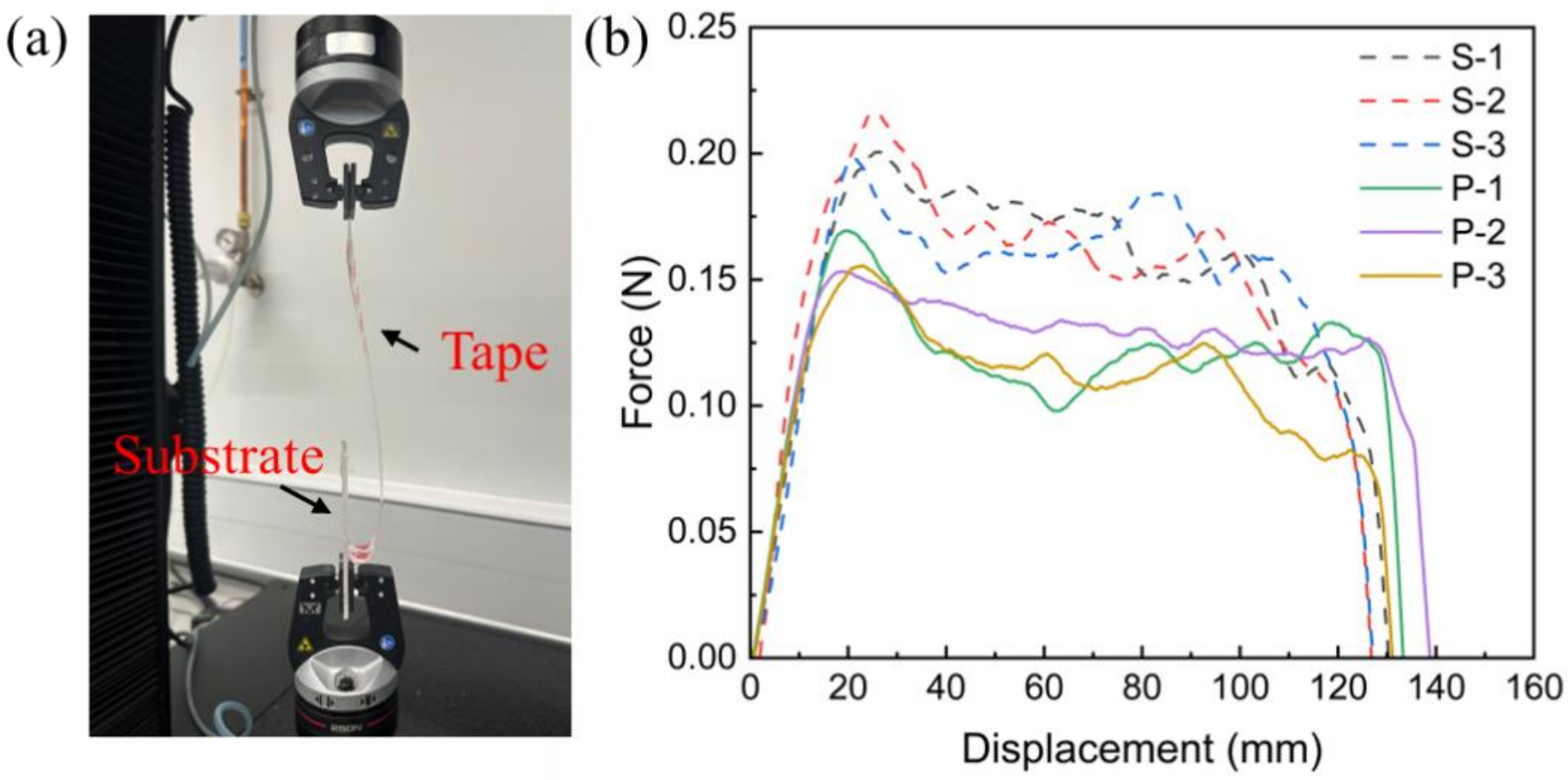

Figure S5. (a) Experimental setup for the 180° peeling test; (b) Load-displacement curves for tapes peeling from pristine PDMS substrate (P-1, P-2 and P-3) and speckled PDMS substrates (S-1, S-2 and S-3).

The peeling tests revealed that the speckled PDMS exhibited a higher steady-state peeling force of 0.175 N ± 0.015 N compared to 0.125 ± 0.018 N for pristine PDMS, corresponding to energy release rates of approximately 7.0 ± 0.6 J $m^{-2}$ and 5.0 ± 0.7 J $m^{-2}$, respectively. The displacement curves of the speckled PDMS displayed more pronounced fluctuations and earlier failure initiation, suggesting a heterogeneous interfacial structure. These results imply that while the surface chemistry remained largely unchanged, the introduction of pigment particles modified the microscale topography and local mechanical properties, leading to enhanced mechanical interlocking and crack-path deflection that increased overall adhesion but reduced interfacial uniformity and stability during peeling.

**S.5 Validation of Winkler foundation via FE simulation**

FE simulations were conducted using a 3D elastic substate illustrated in Figure S6a (Young's modulus E = 1.7 MPa, Poisson's ratio ν = 0.30–0.45, thickness h = 0.3 mm) with a total surface area of 22.5 × 22.5 $mm^2$. The central region (7.5 × 7.5 $mm^2$) was subjected to uniform surface pressures in the z direction ranging from 0.05 and 0.25 MPa. The mesh type was C3D8R, and the mesh density was approximately 475 elements/$mm^3$, consistent with the DIC resolution. The nodes at the bottom surface were fully constrained to prevent any translational or rotational movement.

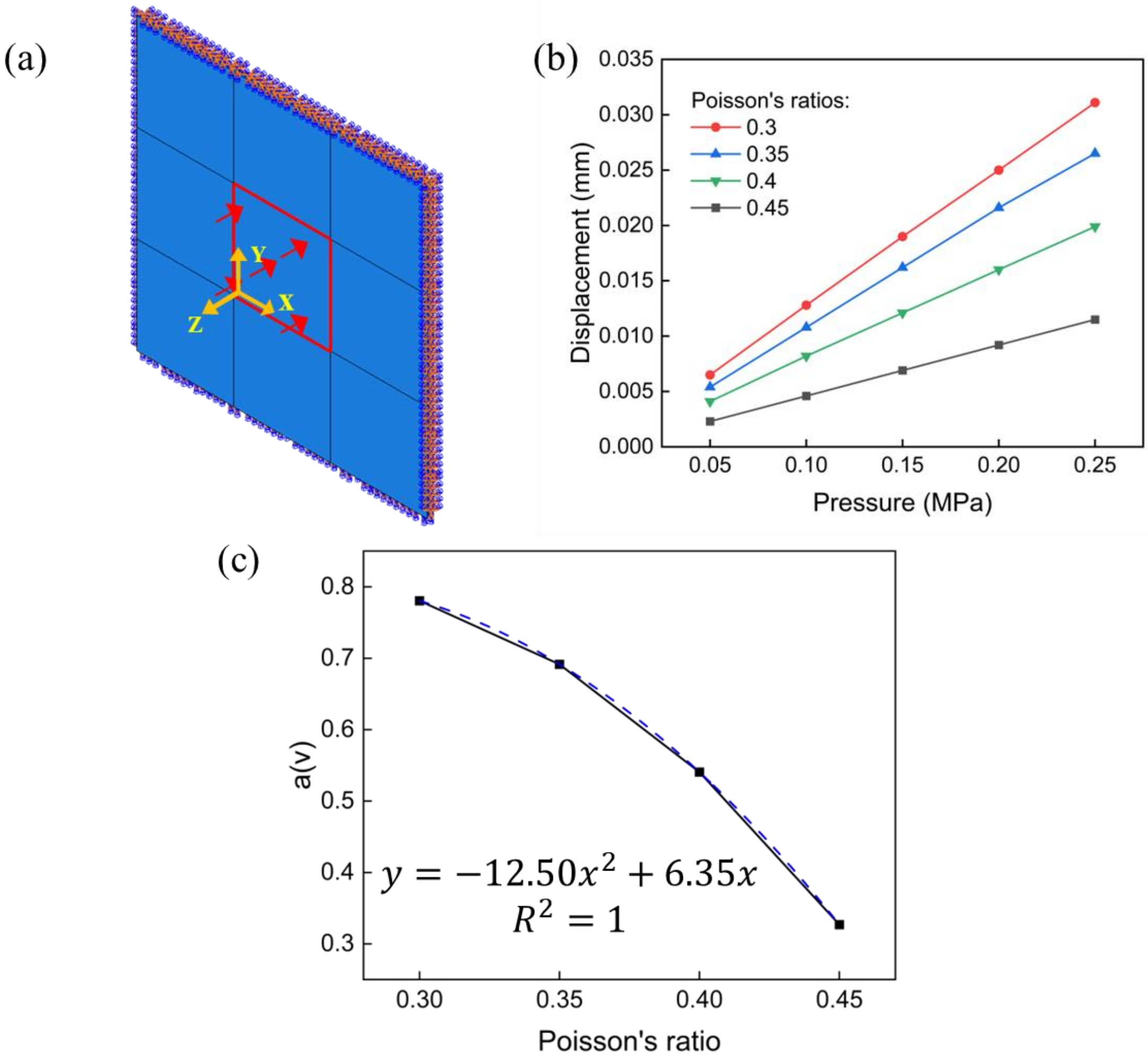

Figure S6. (a) The FE model used for validation of the Winkler foundation approximation;(b) The displacement-pressure curves of the substate with Poisson's ratios varying from 0.30 to 0.45); (c) Fitting curve showing the relationship between the modifying parameter *a(v)* and Poisson's ratio

The resulting surface displacements varied linearly with applied pressure, consistent with linear-elastic response, see Figure S6b. The effective stiffness values obtained from the simulations, $k_{FE} = {}^{\sigma}/_{\delta_z}$, for different Poisson's ratios are summarised in Table S3. The corresponding stiffness values calculated using Timoshenko solution, $k_T = \frac{E}{h}(1 - v^2)$ are also included for comparison. The results show that both the $k_{FE}$ and the $k_T$ increase with Poisson's ratio. However, $k_{FE}$ exhibits more significant dependency, increasing by up to 172%, compared to only 14% for $k_T$.

Table S3. Effective stiffness $k_{FE}$ obtained from simulations for different Poisson's ratios, compared with analytical stiffness $k_T$ from the Timoshenko solution, and the corresponding modifying parameter *a(v)*.

| **Poisson' ratio** | **$k_{FE}$ (N/mm$^3$)** | **$k_T$ (N/mm$^3$)** | ***a(v)*** |
|---|---|---|---|
| 0.30 | 7.98 | 6.23 | 0.78 |
| 0.35 | 9.34 | 6.46 | 0.69 |
| 0.40 | 12.48 | 6.75 | 0.54 |
| 0.45 | 21.74 | 7.11 | 0.33 |

A modifying parameter, $a(v) = \frac{k_T}{k_{FE}}$ was introduced to quantify the deviation of the FE-predicted stiffness from the classical Winkler-based approximation due to lateral constraint and finite-thickness effects. The results are summarised in Table S3. Polynomial fitting of *α(v)* as a function of Poisson's ratio shown in Figure S6c yielded the following relationship:

$$a(v) = (-12.50 \times v^2 + 6.35 \times v) \tag{1}$$

The fit exhibited an excellent correlation ($R^2 = 1.0$), indicating that the polynomial expression accurately reproduces the dependence of effective foundation stiffness on Poisson's ratio. To incorporate both the Poisson-dependent term $(1 - v^2)$ from the Timoshenko solution and the fitted correction factor $a(v)$ into a single function, they were combined into a unified coefficient $\theta(v)$, expressed as:

$$\theta(v) = a(v)(1 - v^2) \tag{2}$$

Incorporating this correction into the Winkler foundation model gives the modified form:

$$k = \frac{E}{h} \times \theta(v) \tag{3}$$

This formulation provides a Poisson-ratio-dependent expression for the effective foundation stiffness, bridging the simplified Winkler model with the continuum-mechanical response of a finite-thickness elastic layer.

## S.6 Filtering noises in measured displacements

Noise may inevitably contain in the displacement computed by the DIC due to correlation algorithms (*61*) and the vibration of the platform. The noise was assumed to occur in all the areas of the substrate and lead to the rigid body displacements in the undeformed area of the substrate. The left edge of the substrate was assumed to be undeformed area and the displacements in X, Y and Z directions were averaged for all the subsets as below:

$$\overline{N_x} = \frac{\sum_{i=1}^{n} N_{ix}}{n}$$
$$\overline{N_y} = \frac{\sum_{i=1}^{n} N_{iy}}{n} \quad (2)$$
$$\overline{N_z} = \frac{\sum_{i=1}^{n} N_{iz}}{n}$$

where $N_{ix}$, $N_{iy}$ and $N_{iz}$ denote to the computed displacement of the $i_{th}$ subsets along the left edge of the substrate in X, Y and Z direction, respectively. $n$ denotes to the number of subsets along the left edge. $\overline{N_x}$, $\overline{N_y}$ and $\overline{N_z}$ denote to the average displacement in the X, Y and Z direction, respectively.

The noise contained in the computed displacement of subsets was alleviated by reducing the $\overline{N_x}$, $\overline{N_y}$ and $\overline{N_z}$ as below:

$$D_x = D'_x - \overline{N_x}$$
$$D_y = D'_y - \overline{N_y} \quad (3)$$
$$D_z = D'_z - \overline{N_z}$$

where $D'_x$, $D'_y$ and $D'_z$ denote to the computed displacements of subsets in X, Y and Z direction, respectively. $D_x$, $D_y$ and $D_z$ denote to the displacement of subsets after alleviating the noise in X, Y and Z direction, respectively. The $D_x$, $D_y$ and $D_z$ were used as the 3D displacement vector which was applied on nodal points of the top surface of the PDMS substrate via FE.

As illustrated in Figures S7 through Figures S10, this filtering operation effectively translates the displacement fields in the X, Y, and Z directions to correct the baseline offset caused by systematic noise. This correction prevents the overestimation of displacement and traction force magnitudes. This effect is particularly evident at the boundaries of the measurement area, where the theoretical displacement and force should be zero. The plots confirm that the filtering process successfully shifts these boundary values down to near-zero levels, ensuring physical consistency.

To quantitatively evaluate the noise reduction efficacy, we analyzed the residual signals at the deformation boundaries (the "remained noise") for both the control experiment and the biological application. In the steel ball case (Figures S7 and S8), the unfiltered displacement magnitude at the deformation boundary reached approximately 3.5 µm. Post-filtering, the magnitude was effectively reduced to the sub-micron level (approximately 0.5 µm). Specifically, the residual noise induced upper and lower bounds of [0.4, 1.2] µm (Figure S7b) and [0.4, 1.1] µm (Figure S8b) (within 4% of the peak displacement). Similarly, the traction force magnitude at the boundary decreased from a peak of 1.5 mN (noise-contained) to a residual level of approximately 0.2 mN, bounded by [0, 0.4] mN (within 5% of the peak load). Based on these results, the traction force uncertainty boundary is estimated to be approximately ±0.4 mN.

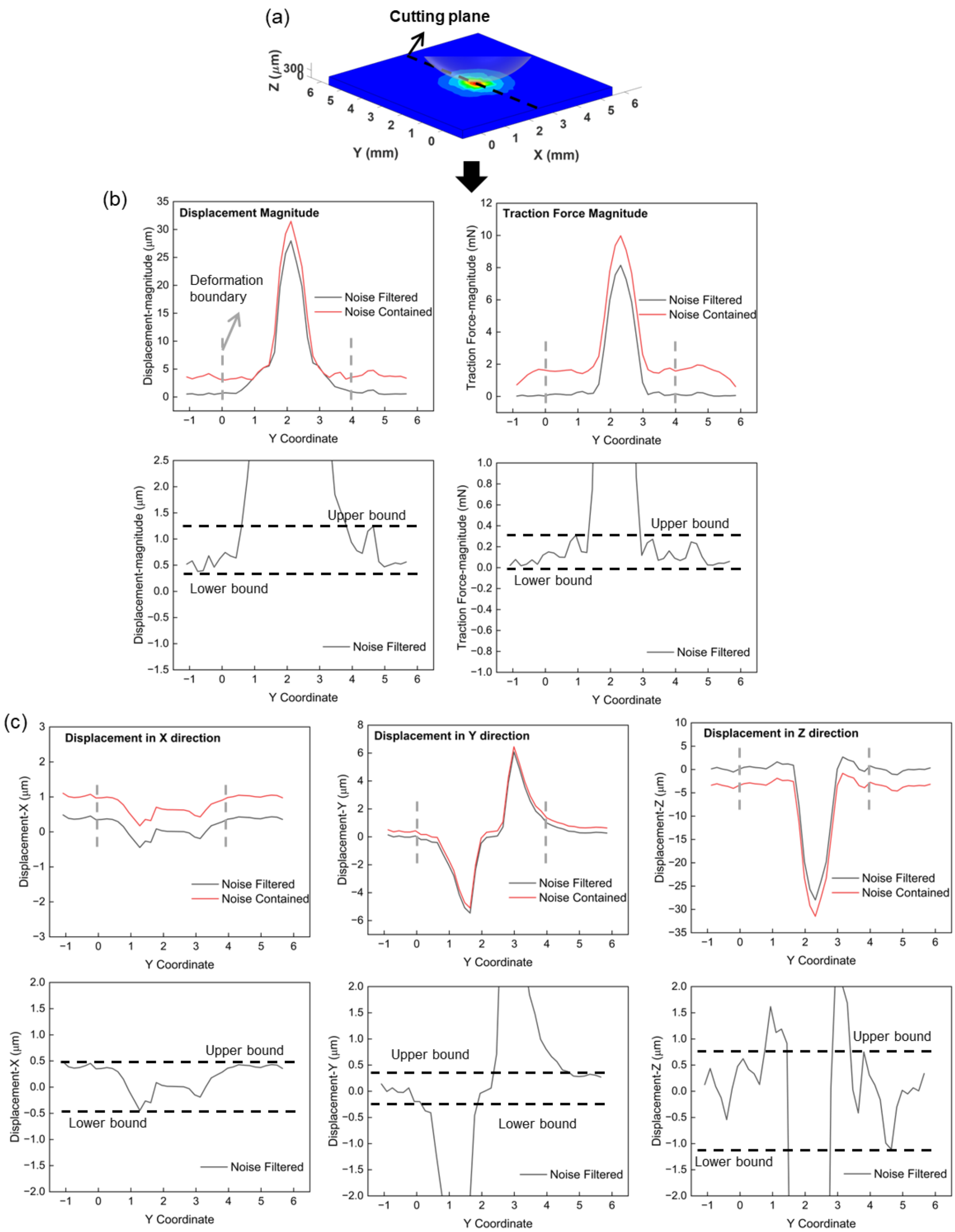

Figure S7. Steel ball under dry condition: Uncertainty quantification along the Y-axis cut. (a) Illustrative cutting view along the Y-axis at the central X-axis plane. (b) Comparison of noise-filtered and noise-contained magnitudes. The residual noise (after denoising) induces upper and lower bounds of [0.4, 1.2] μm for displacement magnitude and [0, 0.3] mN for traction force magnitude. (c) X/Y/Z displacement fields showing residual noise bounds of [-0.5, 0.5] μm, [-0.3, 0.3] μm, and [-1.1, 0.7] μm, respectively.

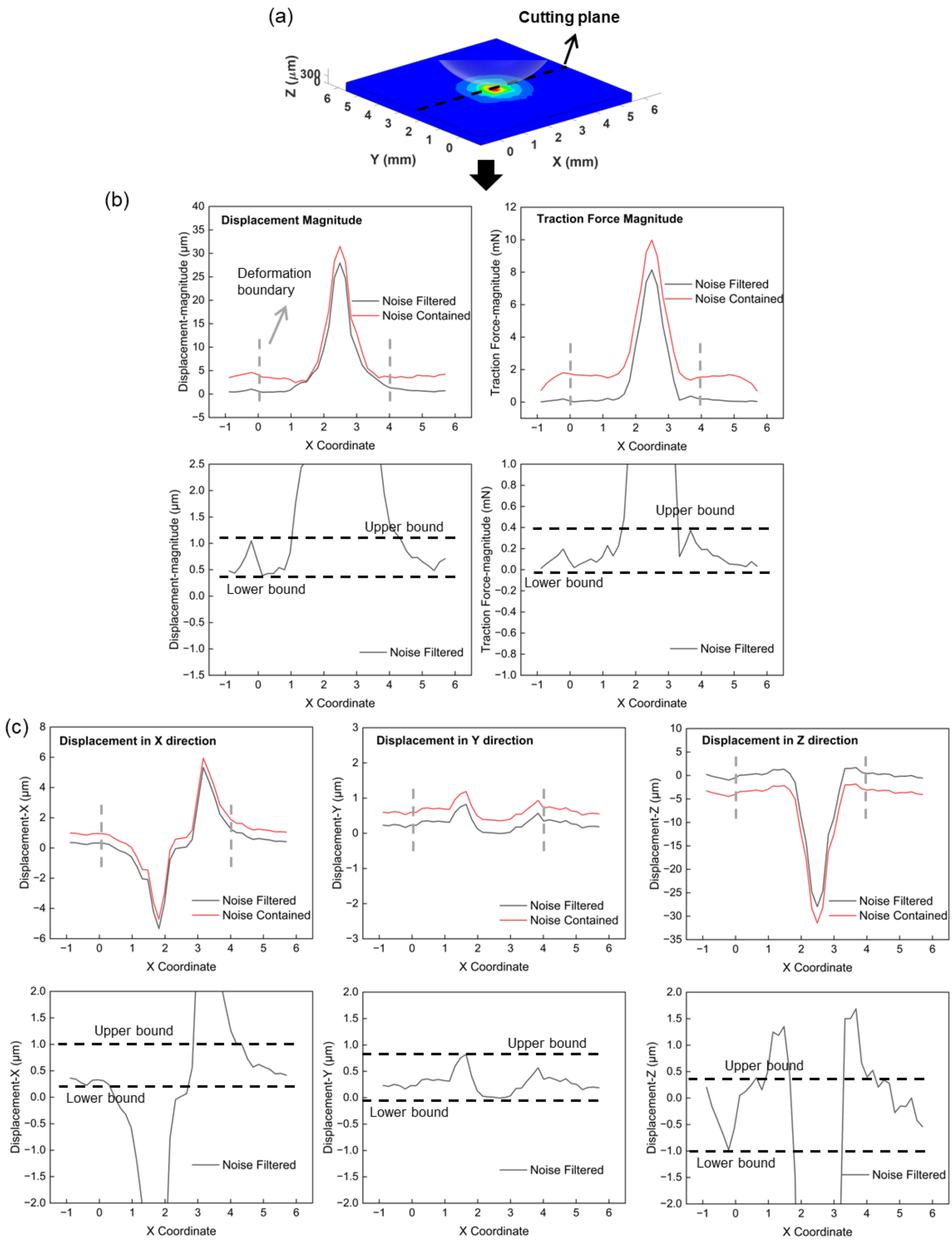

Figure S8. Steel ball under dry condition: Uncertainty quantification along the X-axis cut. (a) Illustrative cutting view along the X-axis at the central Y-axis plane. (b) Comparison of magnitudes showing residual noise induced bounds of [0.4, 1.1] µm (displacement) and [0, 0.4] mN (traction force). (c) X/Y/Z displacement fields showing residual noise bounds of [0.2, 1.0] µm, [-0.1, 0.8] µm, and [-1.0, 0.4] µm, respectively.

A similar trend was observed in the mussel plaque case (Figures S9 and S10). The displacement magnitude at the deformation boundary dropped from about 4.2 μm (noise-contained) to approximately 0.3 μm (noise-filtered). The residual noise bounds for displacement components were narrowed to generally within [-0.4, 0.4] μm (within 1% of the peak displacement). Consequently, the traction force magnitude was reduced from 1.3 mN to a residual level close to 0 mN, bounded by [0, 0.2] mN (within 5% of the peak load). Therefore, the traction force uncertainty boundary for the bio-adhesion experiments is estimated to be approximately ±0.2 mN. It was noted in the cut view along the X-axis (at the central Y-axis plane) that the noise-filtered and noise-contained traction force curves intersect at the boundary region. This suggests that while the overall systematic noise is suppressed, the traction forces at the exact periphery might still be slightly influenced by the filtering process. Future work will focus on refining the algorithm to further minimize these boundary artifacts and improve edge precision.

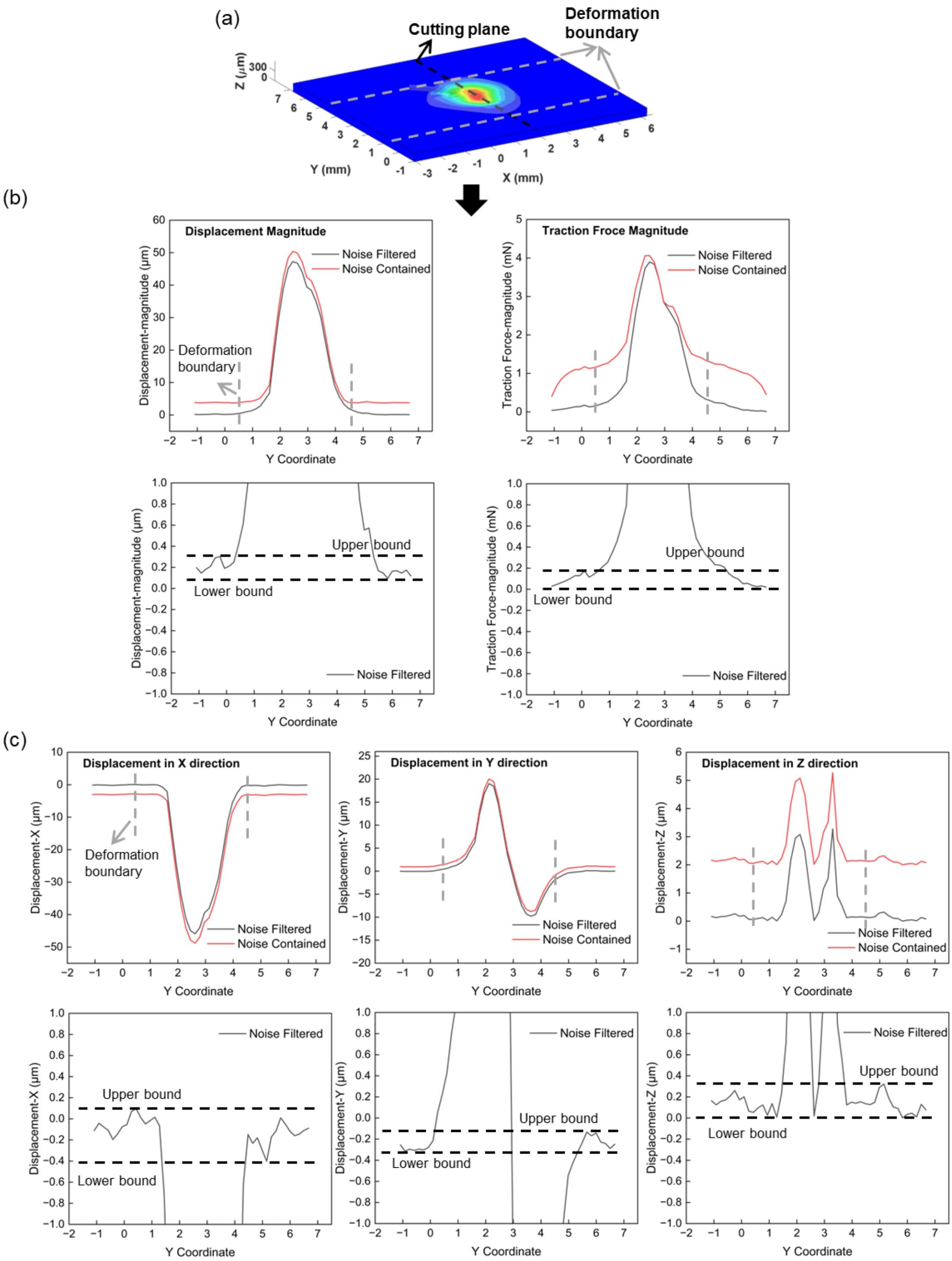

Figure S9. Mussel plaque under 15° tension: Uncertainty quantification along the Y-axis cut. (a) Illustrative cutting view along the Y-axis at the central X-axis plane. (b) Comparison of magnitudes showing residual noise induced bounds of [0.1, 0.3] μm (displacement) and [0, 0.2] mN (traction force). (c) X/Y/Z displacement fields showing residual noise bounds of [-0.4, 0.1] μm, [-0.3, -0.1] μm, and [0, 0.3] μm, respectively.

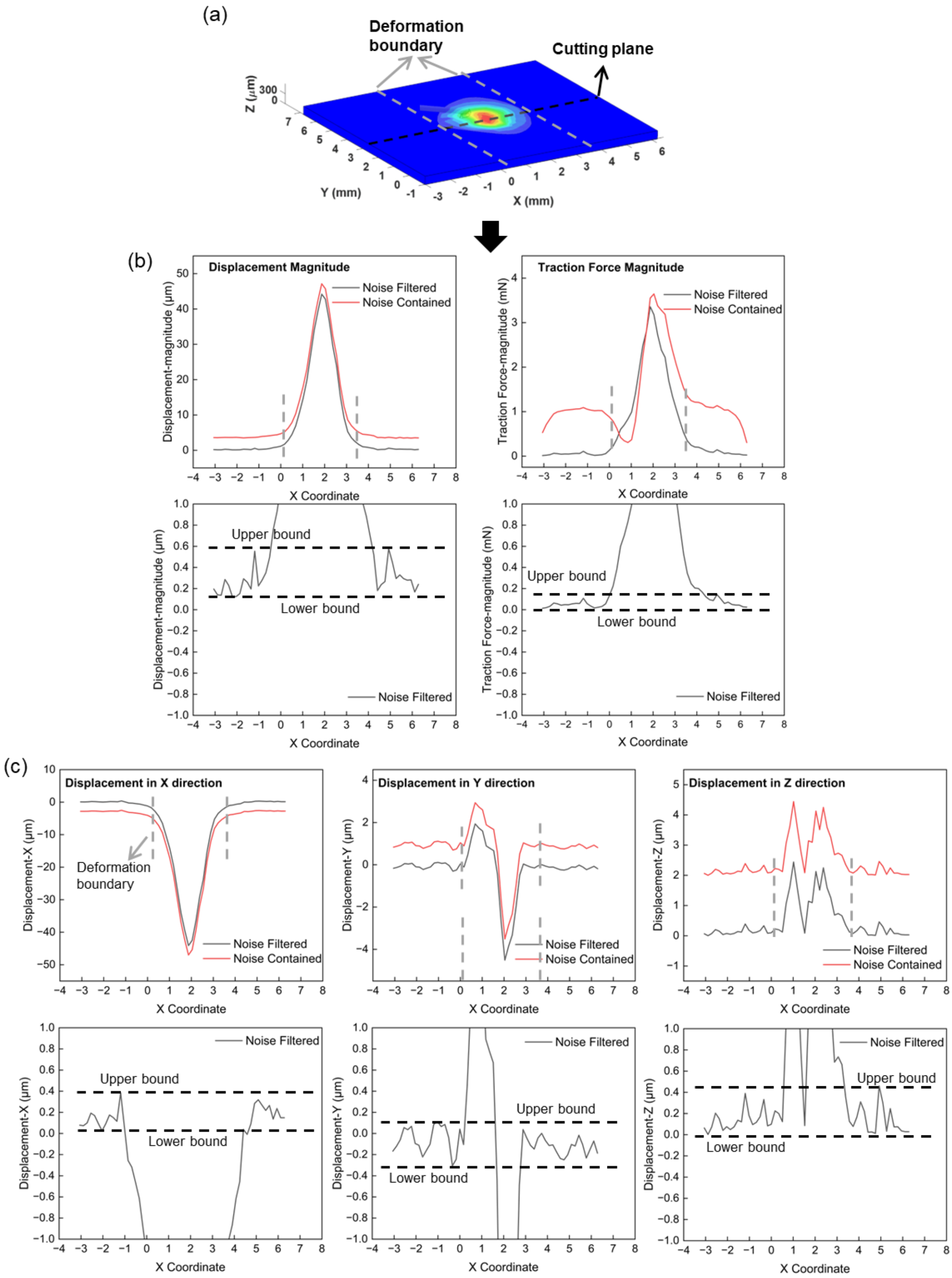

Figure S10. Mussel plaque under 15° tension: Uncertainty quantification along the X-axis cut. (a) Illustrative cutting view along the X-axis at the central Y-axis plane. (b) Comparison of magnitudes showing residual noise induced bounds of [0.1, 0.6] μm (displacement) and [0, 0.2] mN (traction force). (c) X/Y/Z displacement fields showing residual noise bounds of [0, 0.4] μm, [-0.3, 0.1] μm, and [0, 0.5] μm, respectively.

**S.7 Measurement deviations**

The method was calibrated to evaluate the effects of DIC parameters on the measurement accuracy. A steel ball (diameter: 13 mm, mass: 9.0 g) was placed on the PDMS substrate in wet condition, and the substrate's deformations in the X, Y, and Z directions were measured. The results were used to calibrate the DIC parameters. The 9 g ball was selected for calibration because its gravity (~0.09 N) is close to the minimum tension force of mussel plaques' detachment on the substate.

In current study, the numbers of subsets within the ROI region for interval $L_i$=10, 15, 25, 35, 45 pixels were 1748, 750, 270, 143, and 90, respectively. After determining the number (N) of subsets, the displacements of all the subsets in the X, Y, and Z directions were measured (Figure S11). To evaluate the influence of interval $L_i$, the mean displacement ($\overline{D_i}$) for each subset $i$ at a given interval was calculated by averaging the displacements of corresponding subsets obtained using four different subset sizes (15, 25, 35, and 45 pixels) as follows:

$$\overline{D_i} = (D_{i1} + D_{i2} + D_{i3} + D_{i4})/4 \quad (5)$$

Here, $D_{i1}$ to $D_{i4}$ denote the displacements of subset $i$, each obtained from displacement fields computed using the four different subset sizes, respectively. The deviation between each measured displacement ($D_i$) and the mean displacement was determined using the root mean square deviation (RMSD) as follows:

$$RMSD_j = \sqrt{\frac{\sum_{i=1}^{N}(D_i - \overline{D_i})^2}{N}} \quad (6)$$

where $j$ denotes the directions X, Y and Z. The deviation relative to the maximum displacement was then calculated as below:

$$\theta = {RMSD_j}/{D_{max}} \quad (7)$$

where $\theta$ is the relative deviation and $D_{max}$ is the absolute value of the maximum displacement in X, Y and Z directions. The diameter of deformed area is illustrated in Figure S11 (c), which is approximately 2.5 mm.

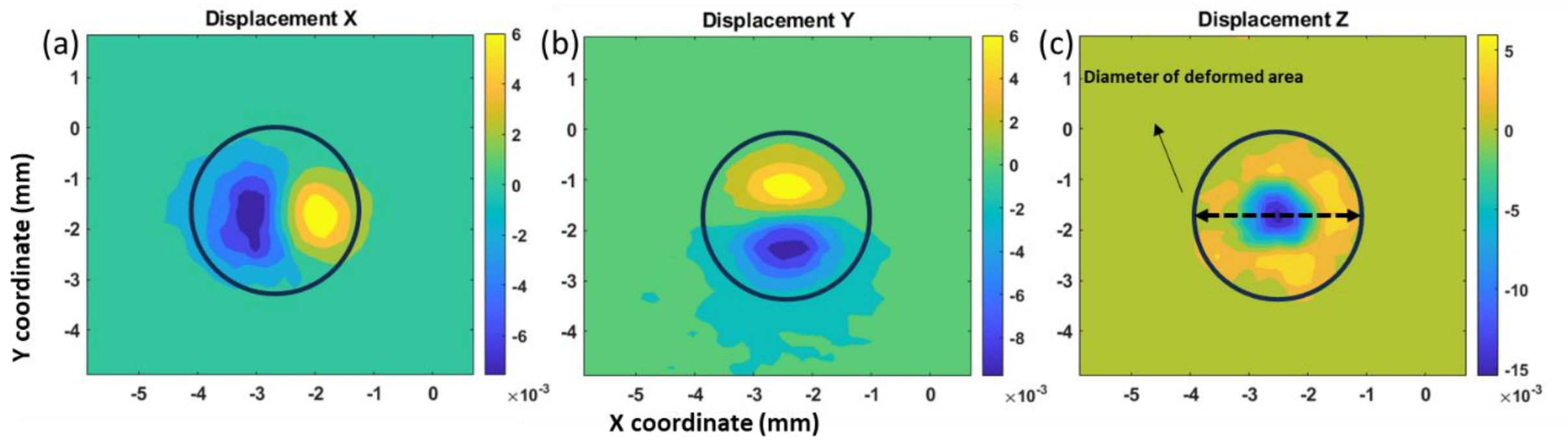

Figure S11. The measured deformations of the substrate under the 9g steel ball in X (a), Y(b) and Z(c) directions.

The distance between adjacent square subsets ($x_i$, $y_j$) and ($x_i$, $y_{j-1}$), was determined by calculating the absolute difference of their y-coordinates ($|y_i - y_{j-1}|$). Since the subsets are square, the same calculation applies to the x-coordinates. Figure S12a shows the actual distance (unit: mm) increases linearly with subset interval (unit: pixel). The slope indicates the actual length per unit pixel, i.e., 0.015 mm. Therefore, the subset intervals of 10, 15, 25, 35 and 45 pixels correspond to 0.15, 0.225, 0.375, 0.525 and 0.675 mm, respectively. Similarly, the subset sizes of 15, 25, 35, 45 pixels corresponds to 0.225, 0.375, 0.525 and 0.675 mm, respectively. The subset sizes and intervals are then normalised relative to the ROI of the substrate under 9 g steel ball as below:

$$L_{ns} = {}^{L_s}/_{d} \qquad (3)$$
$$L_{ni} = {}^{L_i}/_{d}$$

where $L_{ns}$ and $L_{ni}$ denote to the normalised subset size and interval, respectively. $d$ denotes to the above-mentioned diameter of the ROI area, i.e., 2.5 mm. Therefore, $L_{ns}$ and $L_{ni}$ are in the range of 0.06 to 0.27 and 0.09 to 20.27, respectively. The absolute values of the measured maximum displacement in X, Y and Z directions are about 6, 8, 15 µm, respectively. The $\theta$ values for the different normalized sizes and intervals are shown in Figure S12 (b-d).

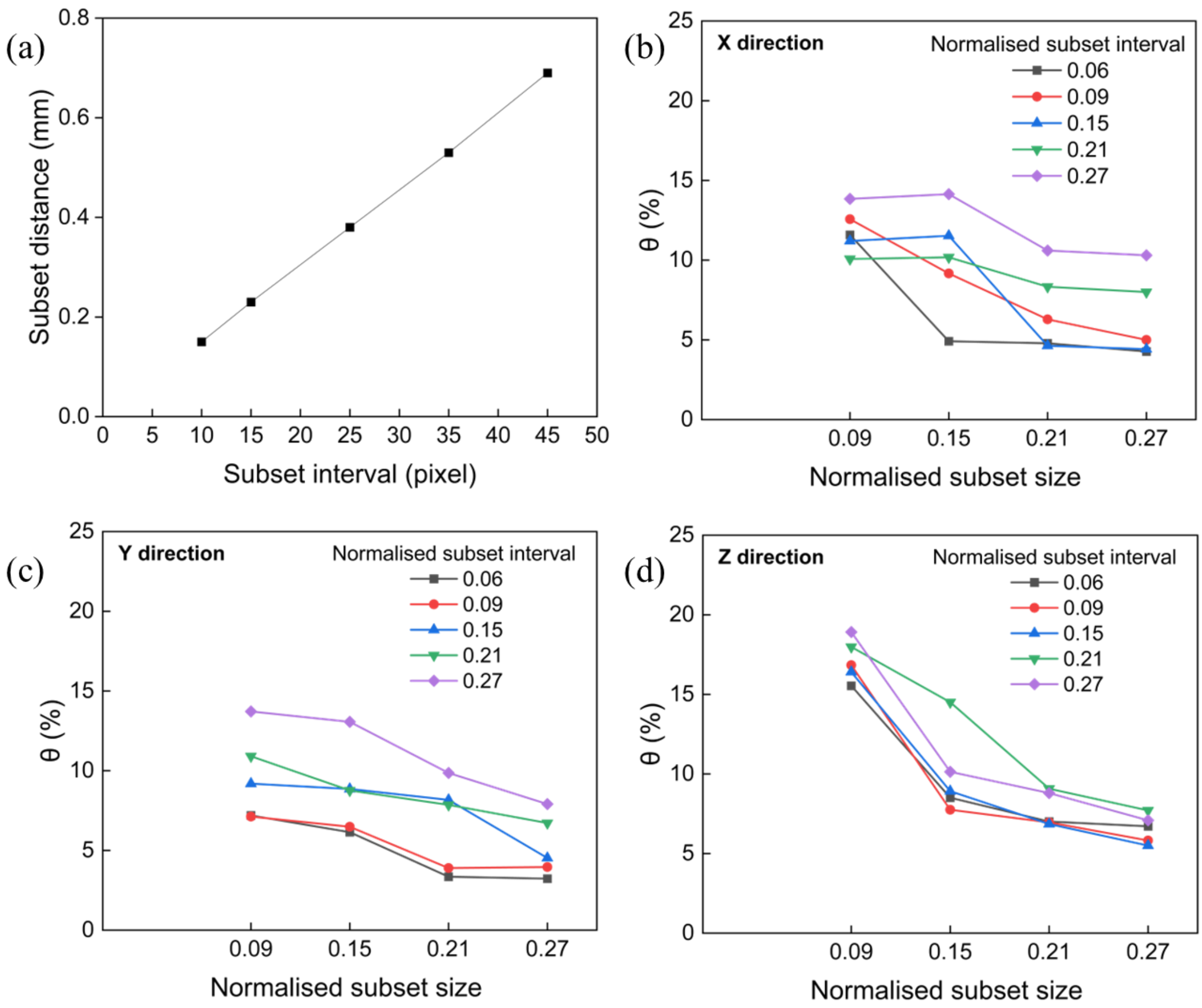

Figure S12. (a) The relationship between subset actual distance in global coordinate and the pixels in images; The deviation relative to deformation as a function of normalised size for each subset interval in X (b), Y (c) and Z (d) direction.

The results show that the displacement in Z direction has the largest $\theta$ (up to about 20%) compared to X and Y directions. The vibration on the platform is mainly vertical which leads to greater noises (up to 2 μm) in the measured Z displacement than noise (less than 1 μm) in X and Y displacements. The $\theta$ values of displacement in the three directions all decrease with the normalised sizes to the minimum of about 5%. This indicates the deviation of the measurement converges for the greater normalised size which comprises enough distinctive speckle pattern for a reliable DIC computation on the displacement.

The subset size should be kept small, as the overall displacement of the entire subset is calculated together, and smaller deformations within the subset may be missed by DIC. This drawback can be alleviated by choosing overlapped subsets when the subset interval is smaller than the subset size. The results show that the $\theta$ decreases with smaller subset intervals, indicating the measurement is more reliable for smaller subset intervals. The results suggest that the reliability of the measurements depend on the size of the subsets which should comprise enough distinctive speckles to be accurately tracked, meanwhile the interval of subsets should be small to capture the deformation of partial areas within a subset. Therefore, the largest subset size (45 pixels) and smallest subset interval (10 pixels) are used for the following tests, consistent with recommendations in the literature (*62*).

**S.8 Quantitative comparison between FE prediction and experimental measurement for steel ball indentation**

To address the discrepancies between the ideal simulated contact stress fields and the experimental results, and to provide a more thorough validation beyond simple contact radius estimation, a quantitative cross-sectional analysis was conducted. Displacement fields and traction force profiles along the central axes of the contact zone for both FE simulations and experimental measurements were compared.

Figure S13 presents the comparison for the steel ball under dry conditions. First, the displacement magnitude profiles for both the experimental and FE results extend across a spatial range of 0–4 mm, effectively defining the deformation zone. The resulting deformation diameters are 3.8 mm for the experimental measurement and 4.0 mm for the simulation, respectively. Near the deformation boundary, the absolute difference of displacement magnitude between the experimental ($U_{Exp}$) and FE ($U_{FE}$) results was less than 3μm. While this corresponds to a maximum relative deviation, defined as $|U_{Exp} - U_{FE}| / U_{Exp}$, of approximately 33.5%, it is crucial to note that this high percentage arises because the displacement values at the boundary are approaching zero, making the ratio sensitive to small variations. This discrepancy is primarily driven by experimental noise. For instance, the experimental X displacement profile along the central Y axis exhibits wave like fluctuations compared to the smooth, linear profile of the FE prediction. In the central deformation region, where the signal to noise ratio is high, the agreement improves significantly, with the relative deviation decreasing to approximately 3.0%. A similar trend is observed in the traction force magnitude ($F_{Exp}$ and $F_{FE}$ refer to experimental and FE, respectively). The absolute difference decreases from less than 0.7 mN at the deformation boundary to 0.2 mN in the central region. In terms of relative accuracy, this central region demonstrates excellent agreement, corresponding to a deviation defined as $|F_{Exp} - F_{FE}| / F_{Exp}$, of approximately 1.9%.

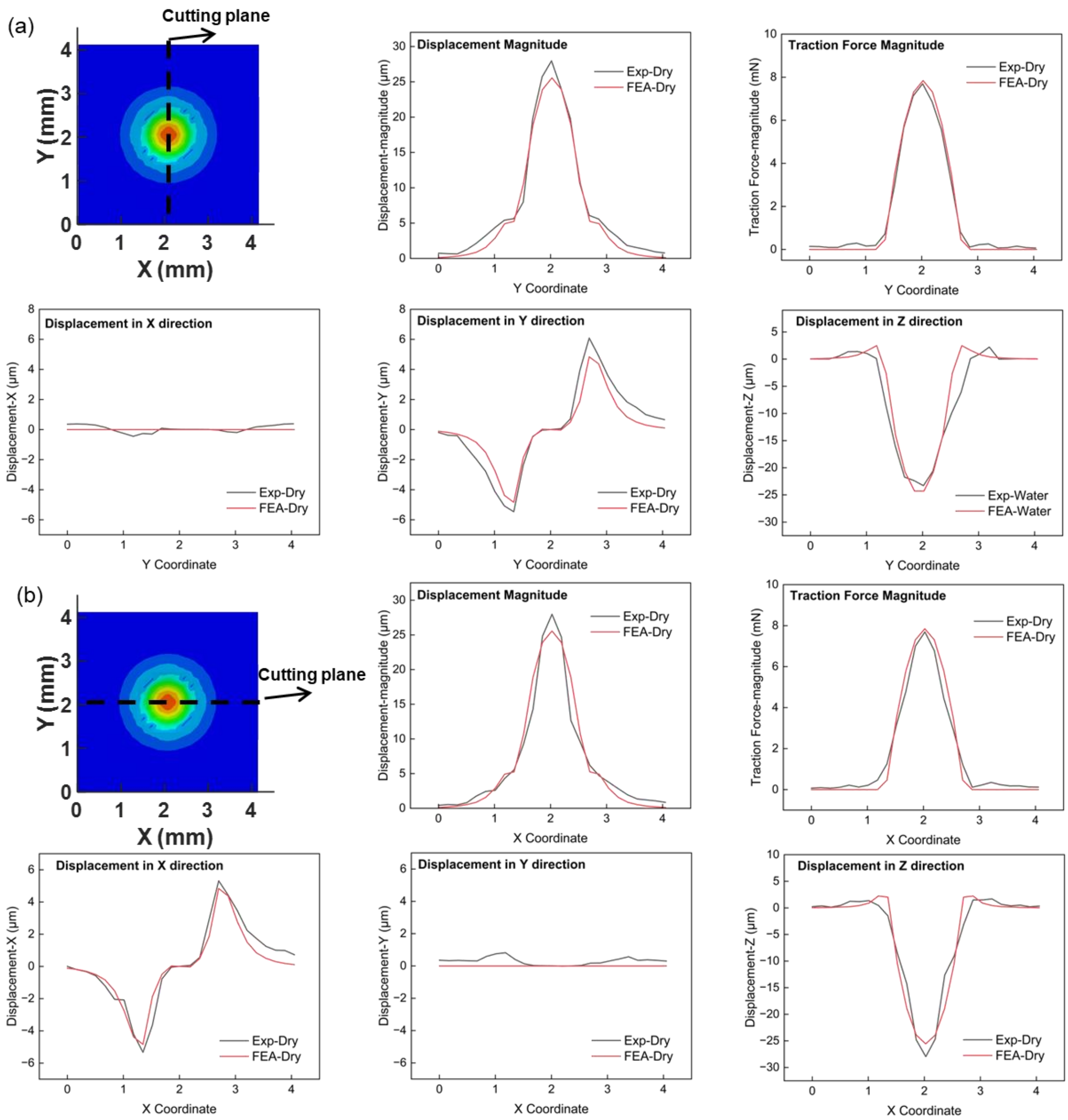

Figure S13. Steel ball under dry condition: (a) FE simulated and experimental measured displacement magnitudes, traction-force magnitudes, and X/Y/Z displacement fields shown on the cut view along the X-axis at the central Y-axis plane. (b) Corresponding fields on the cut view along the Y-axis at the central X-axis plane.

Figure S14 presents the comparison for the wet condition. The absolute difference near the deformation boundary was less than 6 μm. Although this results in a relative deviation of approximately 42.3% due to the low signal magnitude at the edge, the agreement tightens in the central region, where the relative deviation decreases to approximately 3.2%. The absolute difference in traction force magnitude ranged from less than 2.5 mN at the boundary to 0.1 mN in the central region. This central value corresponds to a relative deviation of approximately 1.2%, indicating highly accurate force reconstruction in the primary contact zone. The absolute differences at the deformation boundaries in the wet condition are slightly larger than those in the dry condition. This is attributed to the presence of water, which introduces uncertainties regarding the friction coefficient and its dynamic effect on the contact interface between the steel ball and the substrate, factors that are challenging to model perfectly in FE.

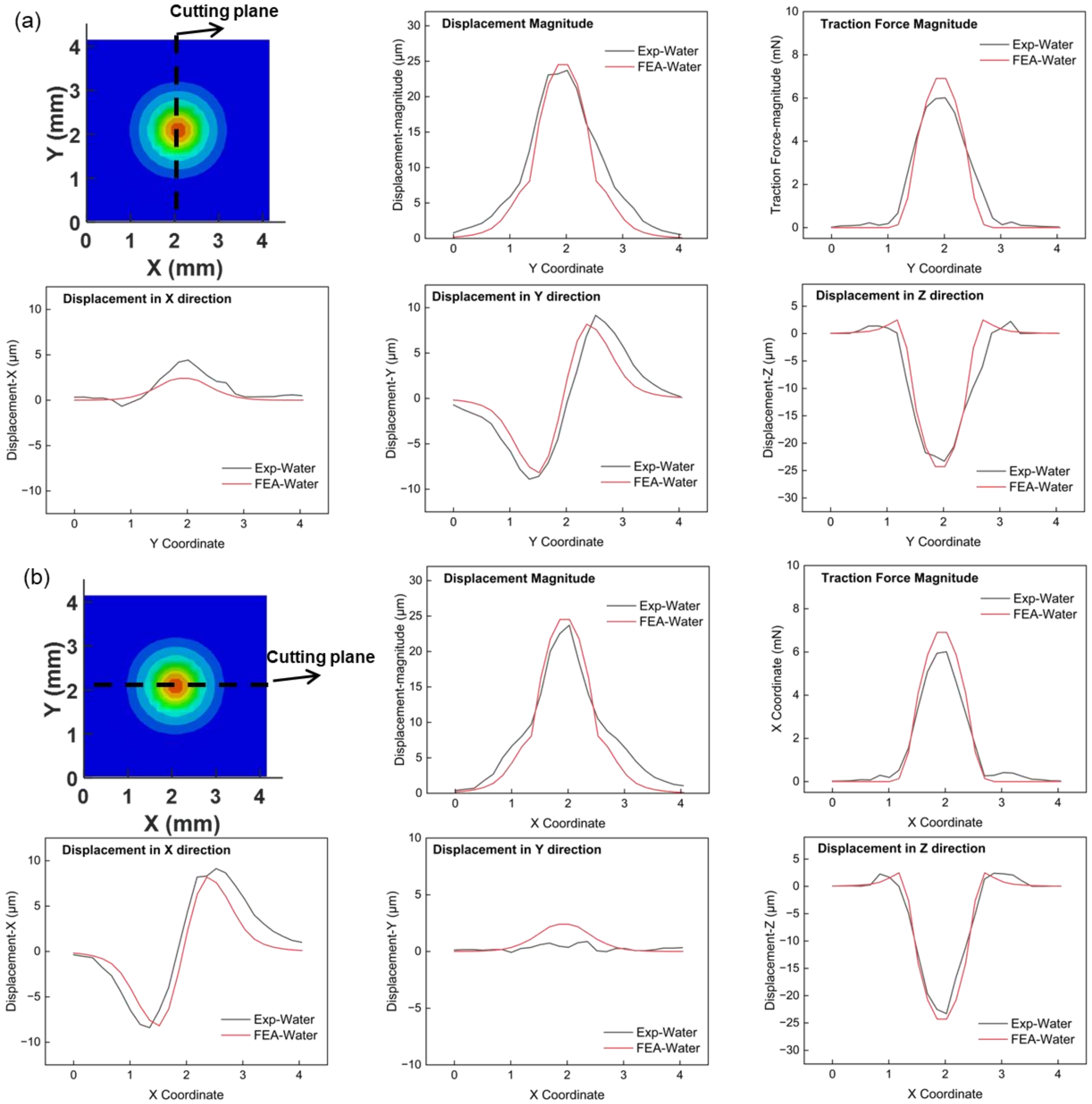

Figure S14. Steel ball under wet condition: (a) FE simulated and experimental measured displacement magnitudes, traction-force magnitudes, and X/Y/Z displacement fields shown on the cut view along the X-axis at the central Y-axis plane. (b) Corresponding fields on the cut view along the Y-axis at the central X-axis plane.

## S.9 Quantitative comparison between linear elastic and hyperelastic models in a controlled simulation

To demonstrate the theoretical necessity of hyperelastic modeling for minimizing measurement error independent of experimental noise, a direct comparison was performed using a purely numerical, well-controlled contact system.

FE simulations were conducted using a 3D substrate model, as illustrated in Figure S15a. The substrate was modeled with a thickness of 0.3 mm and a total surface area of 22.5 mm × 22.5 mm. A central region measuring 7.5 mm × 7.5 mm was subjected to a uniform surface pressure of 2 MPa in the Z-direction. The model utilized C3D8RH elements with a mesh density of approximately 475 elements/mm³, consistent with the resolution of the DIC measurements employed in the main study. The nodes at the bottom surface were fully constrained to prevent any translational or rotational movement. Two distinct material definitions were applied to the substrate: a linear elastic model (Young modulus: 1.38 MPa, Poisson's ratio: 0.45) and a hyperelastic model based on the second-order Ogden parameters derived in Section 2.1. The Young's modulus for the linear model was selected to correspond to the initial stiffness of the fitted Ogden model.

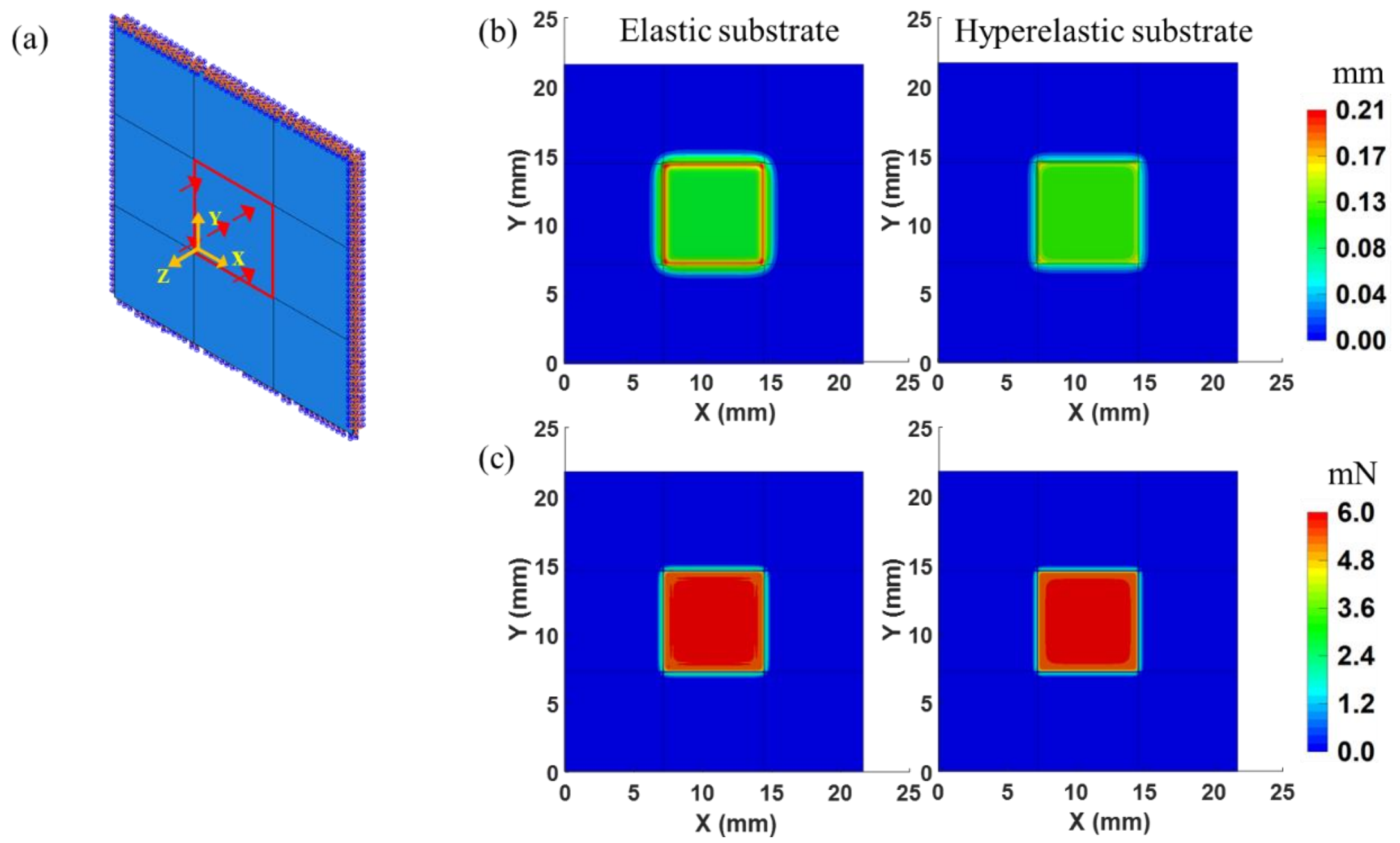

Figure S15. a) Illustration of the 3D substrate model with a central distributed load of 2 MPa; (b) Comparison of displacement magnitude profiles of linear elastic and hyperelastic models; Comparison of reaction force magnitude profiles of linear elastic and hyperelastic models

The comparison reveals that the choice of constitutive model significantly influences both the magnitude and the spatial distribution of deformation and force. As shown in Figure S15b, the displacement profiles for the two models diverge noticeably. In the central loaded region, the displacements exhibit a uniform distribution, forming a plateau with no distinct peaks. The magnitudes for the elastic and hyperelastic models are 0.095 mm and 0.107 mm, respectively, corresponding to a difference of approximately 12.6%. Critically, this deviation is not constant; at the deformation boundary, the difference increases to 19.0%, indicating that the linear model fails to accurately capture the deformation gradient. A similar location-dependent discrepancy was observed in the reaction force distribution (Figure S15c). While the difference between the models is approximately 5.7% in the central region, it widens significantly to 20.0% at the boundary of the loaded zone.

### S.10 The effect of friction between the ball and the substrate

The displacements of the friction (friction coefficient: 1.7) and frictionless models in the X, Y, and Z directions, as well as the overall displacement magnitudes, are summarized in Figure S16. The friction condition does not change circular deformation patterns. However, the diameter (approximately 2.36mm) obtained from friction model was about 7 % lower than that from the frictionless FE simulation (approximately 2.53 mm). Furthermore, the peak values of displacement of the friction model in X, Y Z directions and magnitude are 13%, 13%, 5% and 5% lower than those of the frictionless model, respectively.

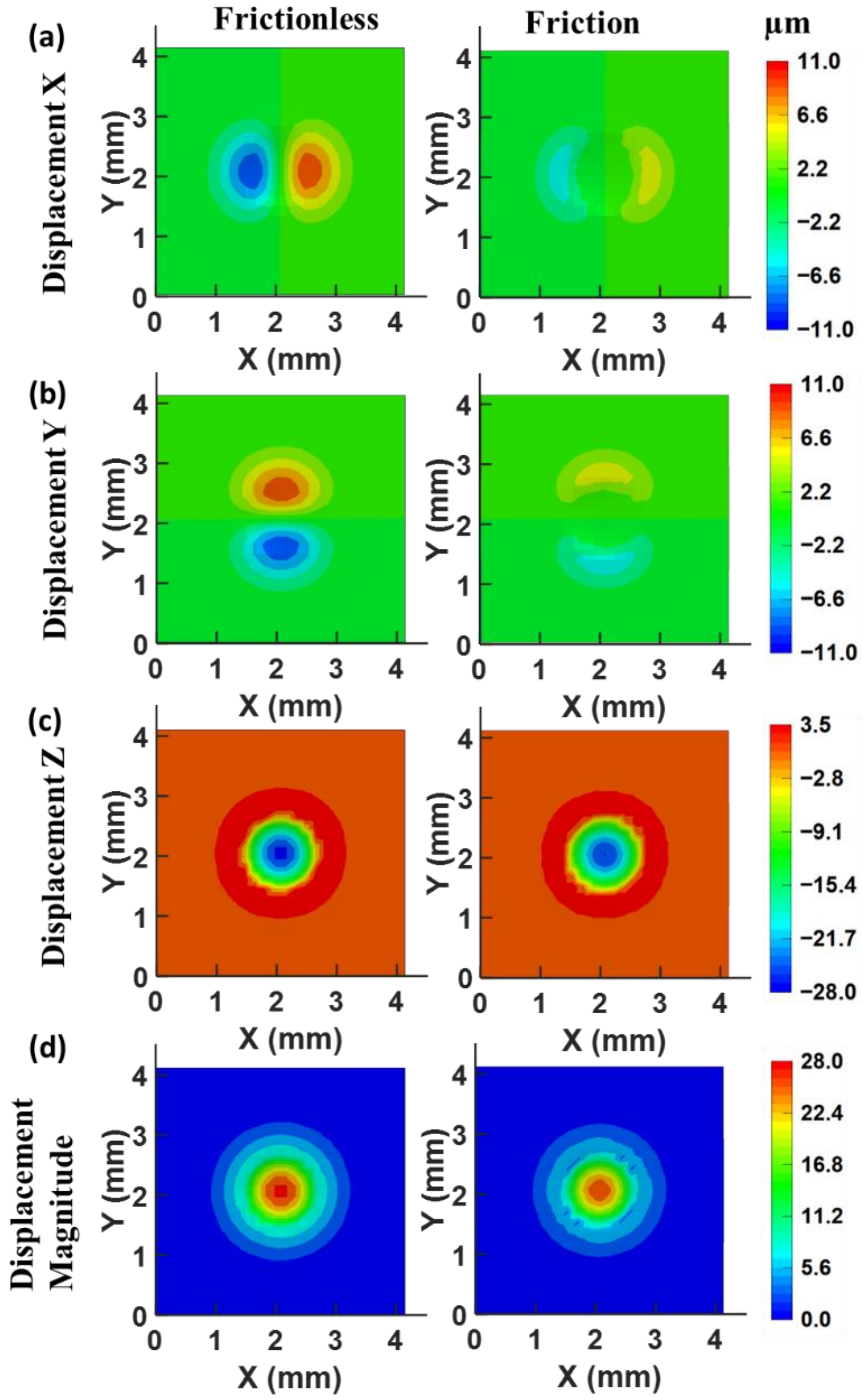

Figure S16. The FE predictions with and without friction conditions of the displacements in (a) X direction, (b) Y direction, (c) Z direction, as well as (d) displacement magnitude.